\documentclass[11pt]{article} 

\usepackage[showcomments]{ajtex}

\usepackage[utf8]{inputenc}
\usepackage{lmodern}

\usepackage{amsthm}

\newtheorem{theorem}{Theorem}
\newtheorem{lemma}[theorem]{Lemma}
\newtheorem{definition}[theorem]{Definition}

\newcommand{\eee}{\mathrm{e}}

\title{Risk aversion can promote cooperation}

\author[1,2,9,10]{Jay Armas}\email{j.armas@uva.nl}
\author[1,2]{Wout Merbis}\email{w.merbis@uva.nl}
\author[1,5,6,7]{Janusz M. Meylahn}\email{j.m.meylahn@utwente.nl}
\author[1,2,4]{Soroush Rafiee Rad}\email{s.rafieerad@uva.nl}
\author[1,3,5,8,11]{Mauricio J. del Razo}\email{m.delrazo@fu-berlin.de}

\affiliation[1]{Dutch Institute for Emergent Phenomena (DIEP), University of Amsterdam}
\affiliation[2]{Institute of Physics (IoP), University of Amsterdam}
\affiliation[3]{van ’t Hoff Institute for Molecular Sciences (HIMS), University of Amsterdam}
\affiliation[4]{Institute for Logic, Language and Computation (ILLC), University of Amsterdam}
\affiliation[5]{Korteweg-de Vries Institute for Mathematics (KdVI), University of Amsterdam}
\affiliation[6]{Informatics Institute (IvI), University of Amsterdam}
\affiliation[7]{Department of Applied Mathematics, University of Twente}
\affiliation[8]{Department of Mathematics and Computer Science, Freie Universität Berlin and Zuse Institute Berlin}
\affiliation[9]{Institute for Advanced Study, University of Amsterdam}
\affiliation[10]{Niels Bohr International Academy, The Niels Bohr Institute, University of Copenhagen \\[2mm]
$^{11}$Zuse Institute Berlin}

\abstract{Cooperative dynamics are central to our understanding of many phenomena in living and complex systems. However, we lack a universal mechanism to explain the emergence of cooperation. We present a novel framework for modelling social dilemma games with an arbitrary number of players by combining reaction networks, methods from quantum mechanics applied to stochastic complex systems, game theory and stochastic simulations of molecular reactions. Using this framework, we propose a novel and robust mechanism for cooperation based on risk aversion that leads to cooperative behaviour in population games. Rather than individuals seeking to maximise payouts in the long run, individuals seek to obtain a minimum set of resources with a given level of confidence and in a limited time span. We show that this mechanism can lead to the emergence of new equilibria in a range of social dilemma games. }

\date{\today}

\begin{document}

\maketitle
\section{Introduction}
\label{sec:introduction}
Cooperation lies at the heart of many phenomena in living and complex systems, including the transition to multicellularity in biology, the emergence of eusociality in insect colonies, and the development of human communities, from ancient hunter-gatherer tribes to modern nation-states \cite{doi:10.1126/science.7466396, axelrod1984evolution}. Understanding the emergence of cooperation between groups and individuals is particularly important for solving global problems such as maintaining ecosystem diversity, mitigating the climate crisis and regulating international financial markets \cite{kaul1999global, 10.1257/aer.15000001, abendschein2021}. However, we still lack knowledge about the possible mechanisms that can lead to large-scale cooperative behaviour in large populations.

Uncovering the mechanisms underlying the emergence and dynamics of cooperation has been the subject of intense study over the past decades, using a variety of methodologies across scientific disciplines, including evolutionary game theory, numerical simulations, and experiments ranging from biology to psychology and behavioural economics. Many of these studies have converged on types of mechanisms that can lead to cooperative behaviour, such as different types of selection (e.g. kin selection, group selection, etc.) and different forms of reciprocity (e.g. direct, indirect, structured, etc.). Depending on the context, these mechanisms can take different forms, such as genetic transmission, cultural transmission, social diversity, reputation, shame, threat of exclusion, sanctions, etc. (see e.g. \cite{10.1111/eva.12303, doi:10.1098/rspb.2008.0829, HAMILTON19641, doi:10.1086/406755, doi:10.1146/annurev-psych-081920-042106, naturesocial, CARPENTER200731, BOYD1989213, spatialchaos, doi:10.1126/science.1133755, doi:10.1073/pnas.1206694109, doi:10.1126/science.1164744, 7e9ca62b-fb0d-33b4-9da4-d2df76b1746a, doi:10.1126/science.1178883}). Others have proposed mechanisms that fall outside this classification, such as the role of commitment explored in \cite{akdeniz2021evolution}.

Most of the studies discussed above have focused on relatively small groups of individuals, for which it has been argued (or shown in specific models) that the cooperative mechanisms uncovered are not scalable to the context of large and diverse groups or populations (see e.g. \cite{Dubreuil2008-DUBSRA, VANVEELEN2009589, BOYD1988337, novakgoods}). It has also been argued that these mechanisms are not general enough, in the sense that they do not explain the emergence of cooperation across species or contexts. We briefly address these arguments. Selection requires individuals or entities to track a common trait, such as the case of kin selection, in which cooperative behaviour is genetically transmitted. Although it may be possible for individuals to track a large group of others with the same trait, cooperation has also been observed between quite different groups of entities (e.g. different types of cells or groups of people). Group selection when dealing with human individuals is typically modelled via cultural transmission and, when focusing on large networks, typically leads to cancellation effects and hence to the absence of cooperation. This casts doubt on the generality of selection as a mechanism for cooperative behaviour. In turn, all forms of reciprocity require not only that there be a sufficient number of interactions between any two individuals for them to learn each other's behaviour, but also that each individual be endowed with an unrealistically large memory so that it can keep track of every single interaction with other individuals. Even if each (human) individual could keep track of all its interactions with a Dunbar number of other individuals, keeping track of a number of individuals several orders of magnitude higher is not feasible, and it is also unlikely that each individual would interact with the same other individual more than once. Moreover, cells in biological environments or insects in ecosystems can only track a few other individuals. This again casts doubt on the generality of reciprocity as a mechanism for the emergence of cooperation.

These considerations suggest the need for new forms of achieving cooperative behaviour. The purpose of this paper is to propose a mechanism that, under certain conditions, can lead to cooperation in any population of at least 3 individuals. To do so, we focus on the prototypical experiments for studying cooperation, namely cooperative games such as the iterated prisoner's dilemma game, which we now briefly describe. In these games, each player faces a social dilemma when interacting with another player: a player can choose to cooperate (C) and pay a cost in order to benefit the other player, or a player can choose to defect (D) and receive benefits at no cost. The idea behind this model is that such games can effectively capture cost-benefit interactions in a wide range of systems, including interactions between cells in a multicellular organism or between insects in a colony. However, in the case of the prisoner's dilemma it is well known that if the game is played for a finite number of $n$ iterations, the outcome (Nash equilibrium) is for all players to defect, and hence no cooperation will emerge. On the other hand, in an infinite game, or when players do not know which iteration the game will end in, cooperation can emerge through various types of reciprocity, as described above, which typically lead players to adopt strategies such as tit-for-tat, expressed by the slogan \emph{if you cooperate, I cooperate and if you defect, I defect} (see e.g. \cite{doi:10.1126/science.7466396, axelrod1984evolution, doi:10.1126/science.1133755}). This setup has been studied in a variety of contexts and different strategies have been found to be optimal for maintaining cooperative behaviour depending on the particular stage of the game. 
\begin{figure}[h!]
    \centering
    \includegraphics[width=0.9\textwidth]{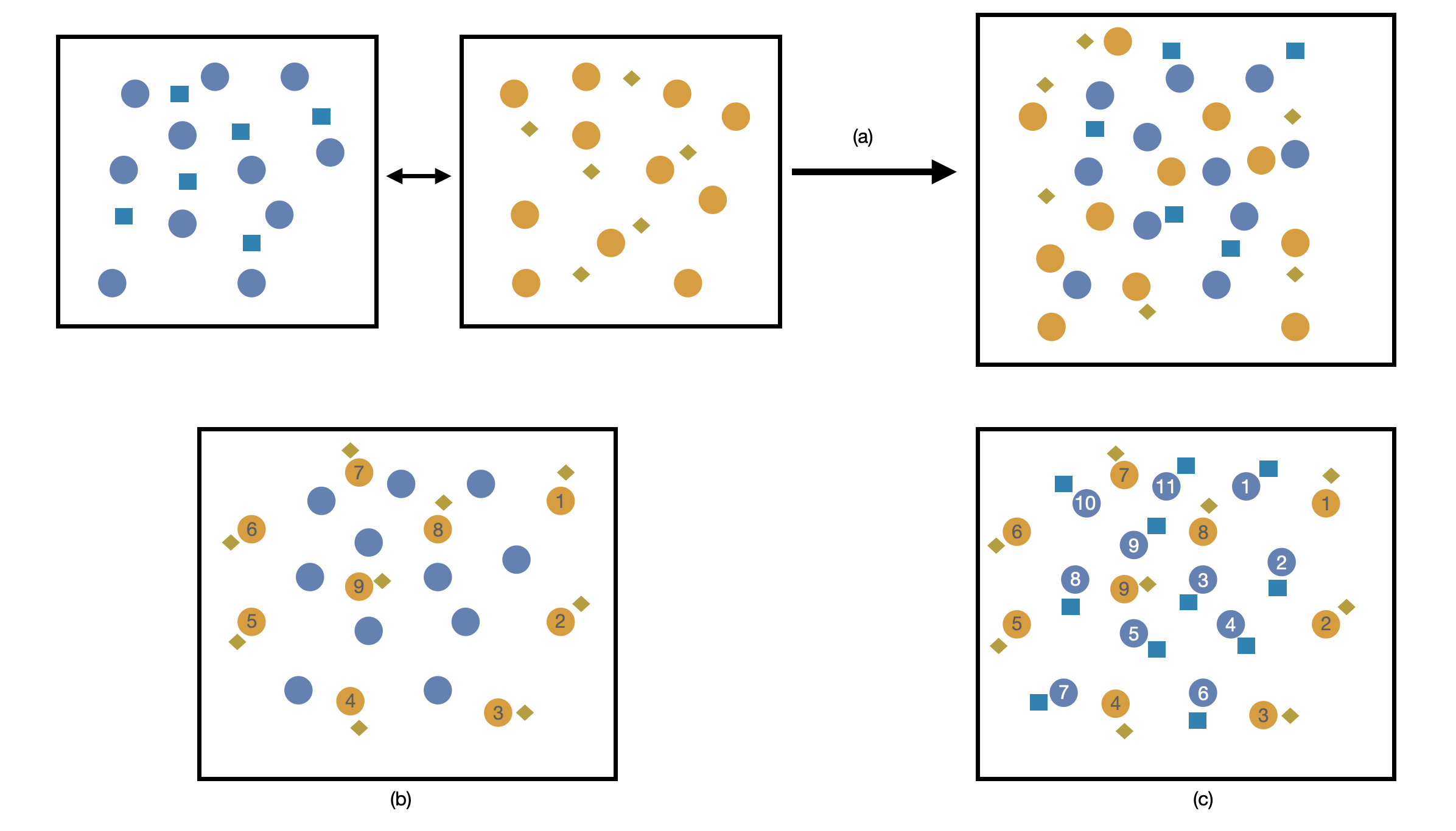}
    \caption{The three different scenarios that we focus on in this paper. In scenario (a) we consider the situation in which a resident population of cooperators (blue circles on the left) who share their resources (blue squares) among themselves is invaded by a population of defectors (yellow circles on the right) who share resources only among defectors. In practice, we consider the situation in which the two populations are homogeneously mixed, as in the top right corner. In scenario (b) we consider the case of a population of cooperators who share their resources among themselves and an invading population of self-driven defectors who do not share their resources. In scenario (c) we consider a control case in which all individuals, cooperators and defectors, are self-driven and do not share resources. }
    \label{fig:scenarios}
\end{figure}

Given the long history of research on cooperation games and specially on the iterated prisoner's dilemma game, it seems pointless to study it further. However, it should be noted that the majority of these results on the emergence of cooperation, as well as those obtained in experiments in psychology and behavioural economics, assume from the outset that each individual is driven by self-interest with the aim of maximising their own fitness (e.g. ability to reproduce in the context of biology), profit, utility or payout in the long run. Although this is a common assumption, it is known that in many cases individuals/entities work for the benefit of the collective. For example, many types of cells work together to maintain the proper functioning of a multicellular organism, bees with different tasks work together to ensure the survival of the hive as a whole, and ants share resources (food) with all other ants in the colony via elaborate transport networks. These are cases where groups share their benefits among group members and try to maximise the collective payout. In these circumstances, it is common to, rather than individual fitness, introduce the notion of group fitness (i.e. the ability of the collective to survive in the long run), which has been argued to emerge in various contexts \cite{multic, 10.1111/evo.14549, doi:10.1073/pnas.0505858102}. From this perspective, collectives are driven towards higher group fitness. 

In this paper, we focus on three situations in which cooperation games are played (see figure \ref{fig:scenarios}) in order to understand the role of group fitness in cooperative dynamics. The scenarios differ with respect to whether benefit accrual for individuals of a specific type is shared or not. In particular, we consider the following scenarios:
\begin{itemize}
\item \textbf{Scenario (a):} a resident population of cooperators who share their benefits and resources among each other (top left square) is invaded by a population of defectors (middle top square) who share their benefits among each other. Since we do not consider spatial dynamics, in practice we are dealing with a homogeneously mixed population (top right square) in which any individual can interact with any other individual. The contexts we have in mind are, for example, a population of resident cells invaded by unicellular (bacterial) organisms; an ant colony invaded by another group of insects; or a group of people (defectors) who do not play fair with another cooperating group in a social or economic context. The same scenario can also model situations where the two groups are actually part of the same collective. For example, the two groups can represent two types of cells interacting in a single multicellular organism.
\item \textbf{Scenario (b):} A resident population of cooperators who share their resources among each other is invaded by several rogue individual defectors who do not share their benefits with other individuals. This scenario can model, for example, an insect colony invaded by other rogue insects, but it can also model traitors or cheaters. Traitors are invaders, but from within the group itself, and are capable of interacting with the group itself, although they aim to maximise their own benefits. Cancer in a multicellular organism would be one example \cite{doi:10.1098/rstb.2014.0219}, while individuals who abandon group participation (e.g. by embezzling funds) would be another. Cheaters, also known as free riders, provide no benefit to the rest of the group. This would be the case of any individual/entity who uses the group for protection but does not contribute to its maintenance, functioning or growth. 
\item \textbf{Scenario (c):} this is a control case against which we compare the other two scenarios. In this scenario, all individuals, either cooperators or defectors, try to maximise their individual fitness and do not share their benefits. By studying this case, we can better understand the effects of maximising group fitness/group utility. 
\end{itemize}

%Paragraph relating our set-up to the usual set-up
Our set-up thus differs from the canonical set-up as described, for example, by \cite{sandholm2010population, nowak2006evolutionary}. We retain the random matching of two individuals to play the game, but usually the ``cooperate'' action itself corresponds to the act of sharing (part of) one's resources with another individual or the entire population, while the ``defect'' action corresponds to free-riding on the sharing of resources by others. In our case, the individuals obtain some reward depending on the actions taken in the game they played, and this reward can then be shared among individuals of the same type as in Scenarios (a) and (b). Our control scenario (c) corresponds to the situation typically considered in the literature (cf. \cite[Chapter 4]{nowak2006evolutionary}). 

\begin{figure}[h!]
    \centering
    \includegraphics[width=0.8\textwidth]{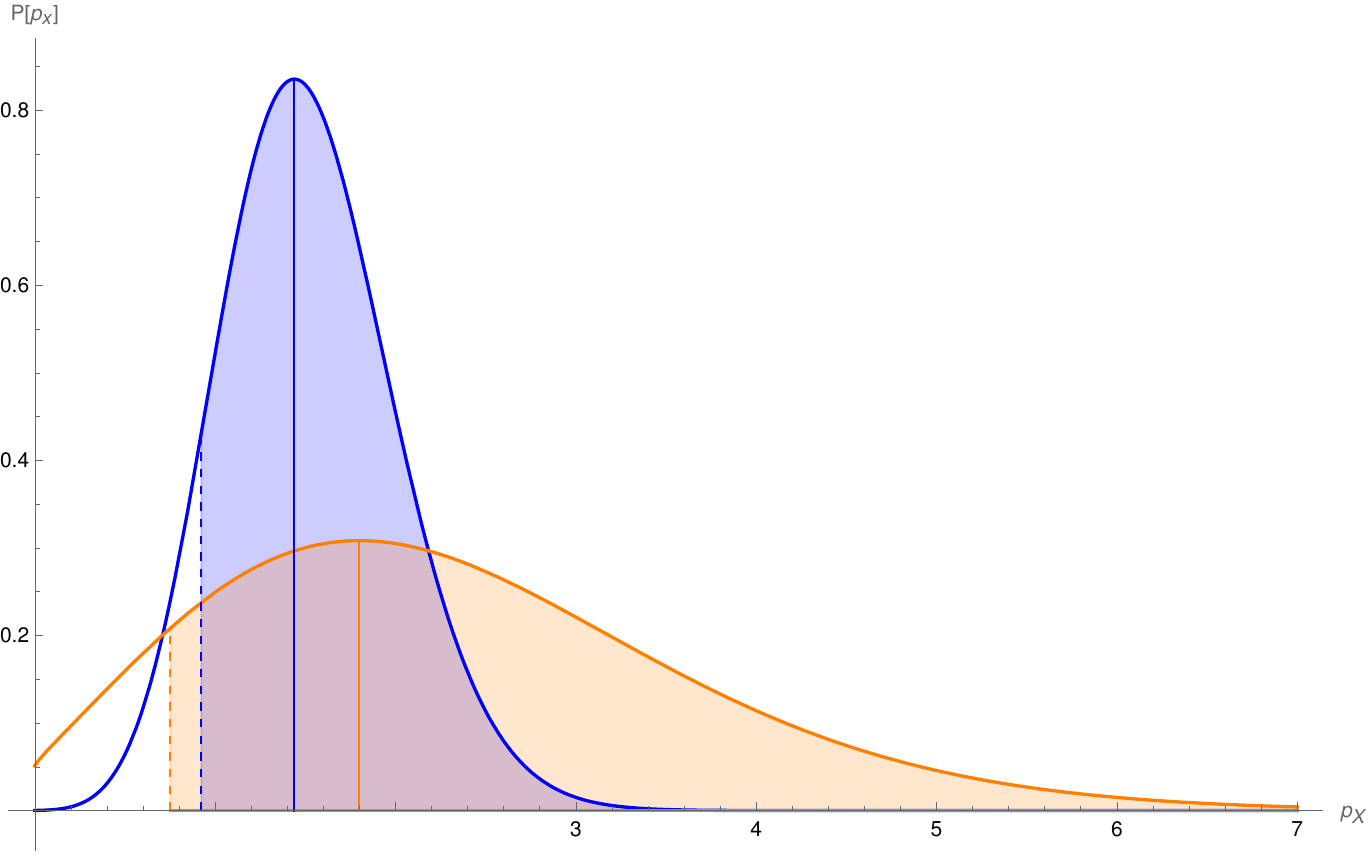}
    \caption{If players base their utility on expected profit rates, they might end up loosing due to stochastic fluctuations in the actually realised profit. risk-averse players might define a different utility based on the minimum profit rates they are sure to receive with a certain fixed probability. In the scenario plotted here, cooperators (blue) have a lower expected profit rate compared to defectors (orange), as indicated by the solid lines. The dashed lines denote the minimum profit rates they can expect to receive with 90\% probability (corresponding to the shaded areas). If risk-averse players base their utility on this quantity, it is beneficial for defectors to switch to cooperating, even though the expected profit rates are higher for defectors. It is this mechanism which enables the emergence of a new Nash equilibrium in populations of prisoner's dilemma players}
    \label{fig:probdistsketch}
\end{figure}
Our goal is to model these different scenarios in detail. However, we will show that if, as is commonly assumed, groups and individuals tend to maximise group or individual utility based on average payouts in the long run, then there is never an incentive for a defector to change their strategy and start cooperating, but only an incentive for cooperators to start defecting. In other words, there is no cooperation. This leads us to ask: is maximising average payout in the infinite future the principle that drives the evolution of cooperation, or even evolution in general? This assumption can be challenged in a number of ways. First, organisms, including humans, live in a finite time, and the timescale for interactions is typically much smaller than their lifespan. In extreme scenarios, for example in scarce or harsh environments, organisms do not seek maximum protection or resources in 50 years' time, instead survival is linked to events that occur in the next few minutes, hours or days, depending on the specific context or stage of development. Second, even if we consider maximising utility in a finite time, rather than the unrealistic infinite future, individuals may be more concerned with ensuring that they receive a minimum payout with a high probability, rather than ensuring that they receive a higher payout with a lower probability. In the context of collective or group survival, maximizing such utility ensures that each individual has the minimum set of resources needed to proceed to the next stage of the game or the next stage of development. 

The general picture we have in mind is shown in Fig.~\ref{fig:probdistsketch}, where we plot the probability $\mathbb{P}(p_X)$ of receiving a payout of $p_X$ for a group of cooperators (in blue) and a group of defectors (in orange). Defectors have a higher average payout $p_X$ than cooperators, represented by solid lines. However, cooperators have a higher minimum payout than defectors with the same probability, represented by the dashed lines. If groups or players are risk-averse, i.e. if they are more likely to aim at receiving a smaller payout but with a high probability\footnote{We use a generalized notion of risk-averseness. The standard definition of risk aversion requires the agents to prefer a certain bet with a lower payout to an uncertain lottery with a higher expected value \cite{Werner2008, aumann2017axiomatic}. We use a generalised notion of risk aversion that requires the agent to prefer an "almost certain" bet to an uncertain one with a higher expected value.}, then there is an incentive for defectors to change their strategy and start cooperating. Throughout this paper, we will show that in scenarios (a) and (b) described above, cooperation emerges even in situations where there are only 2 cooperators and 1 defector. Thus, it is the combination of utility based on risk aversion with group fitness (sharing payouts among members) that leads to cooperative behaviour, as opposed to scenario (c) where each individual may be risk-averse but unwilling to share payouts with a group.  This mechanism appears to be sufficiently robust and scalable from small to very large groups. 

To model games with an arbitrary number of players $N$, we introduce a new framework that combines several methodologies. In particular, we combine evolutionary game theory with chemical reaction networks, as proposed in \cite{veloz2014reaction}, where interactions between players are the result of a \emph{chemical reaction}. However, unlike \cite{veloz2014reaction}, we consider the system to be fully stochastic, where at each time step players are randomly selected from the pool of players to play a given game. It is only for a very large number of players (i.e. in the thermodynamic limit) that the dynamics of our stochastic system collapse to the classical rate equations introduced in \cite{veloz2014reaction} for scenario (a). The mathematical tools developed to study evolutionary game theory are generally drawn from classical probability theory \cite{murray2001mathematical}. Here, instead, we find it useful to use methods developed for many-body quantum mechanics that can be applied to stochastic systems \cite{doi1976second,grassberger1980fock,peliti1985path,schutz1995reaction,Baez_2017,del2024field}.  In this approach, the master equation of the stochastic process (i.e. the Kolmogorov forward equation) can be written in terms of creation and annihilation operators acting on a Fock space representation of the probability vector.  Such methods are useful in describing particle-based reaction-diffusion dynamics \cite{del2022chemical,del2022probabilistic} and have also been used to find analytical solutions for finite-size stochastic systems such as enzyme kinetic models \cite{santos2015fock} and stochastic epidemic models \cite{merbis2022logistic,de2022fock}. The measure of risk aversion that we will introduce relies heavily on these methods. Besides using an analytic stochastic approach in the formulation of cooperation games, we also adapt numerical algorithms suited for bio-molecular reactions \cite{anderson2015stochastic,gillespie1977exact} in order to perform stochastic simulations and obtain information about the system that is hard to obtain analytically. The combination of reaction networks, evolutionary game theory, many-body quantum mechanics tools, and bio-molecular kinetics leads to a novel and robust framework for studying cooperation. In this paper, we use this framework to obtain several promising results on the dynamics of cooperative behaviour. 

We now briefly outline the structure of this paper. In section \ref{sec:basicmodandmachine} we introduce the necessary elements of reaction networks, game theory and stochastic methods that allow us to formulate the stochastic models of cooperation games that we use to study scenarios (a), (b) and (c). In section \ref{sec:Nash} we analyse in detail the Nash equilibrium in these games in two different situations: when players aim to maximise average payouts and in the case of risk aversion. We describe in detail how to define a risk-averse utility function and show that in scenarios (a) and (b) maximising such a function leads to cooperative behaviour. In section \ref{sec:simulation}, we perform stochastic simulations using the Gillespie algorithm and verify as well as improve our analytical analyses. Finally, in section \ref{sec:conclusion} we draw conclusions and discuss possible extensions. We also provide appendix \ref{app:LDT} where we show how to compute the moment generating function for the stochastic models we consider.

\section{Machinery and models}
\label{sec:basicmodandmachine}
The purpose of this section is to introduce the basic elements of game theory, chemical reaction networks and their stochastic formulation, and show how these can be used to formulate models for the different scenarios exposed in figure \ref{fig:scenarios} (for a more pedagogical introduction see \cite{Baez_2017}). In this section we also extract useful information from these models including average profits as well as the moment generating function of the profit distribution for the different groups or rogue individuals which will be crucial for the analysis of Nash equilibria in section \ref{sec:Nash}. 

\subsection{Cooperation games}
\label{sec:cooperationgames}
We consider three well-studied games that model general forms of interaction between individuals that can act as cooperators (C) or defectors (D). If both players cooperate, they both receive the reward $r$. If both players defect, they both receive the punishment $p$ (a smaller gain). If one defects and the other one cooperates, the defecting one receives the temptation payout $\tau$, while the other receives the ``sucker's'' payout, $s$.\footnote{If we set $p=0$ then the various payouts can be interpreted in terms of the cost $\tilde c$ of the action of cooperation and benefit $\tilde b$ of the defection. In particular $r=\tilde b-\tilde c$, $s=-\tilde c$ and $\tau=\tilde b$. } The gains from each combination of decisions are summarised in the table below:
\begin{center}
\begin{tabular}{ c c c }
\hline
   & D & C \\ 
 \hline \hline
 D & $(p,p)$ & $(\tau,s)$ \\  
 C & $(s,\tau)$ & $(r,r)$  \\
\hline
\end{tabular}
\end{center}
The way that the values of $p, r, \tau, s$ relate to each other determines the type of dynamics that cooperation and defecting impose on the interaction of the individuals. The most well-known of these is the prisoner's dilemma (PD) in which the payouts are ordered as $\tau>r>p>s$. In this article, we will also consider two alternative orderings. First, we have $r>\tau>p>s$, which corresponds to the stag hunt (SH) game first described by Jean-Jacques Rousseau in \cite{rousseau1754discourse}. Second, we have $\tau>r>s>p$, which is referred to as the hawk-dove (HD) game in the evolutionary game theory literature, first formulated in \cite{smith1973logic}. For motivational examples of these games, see \cite{rose2010game}. In all three games, $r>p$ ensures that cooperating agents fair better than defecting ones. 

It is possible to model even more general interactions between individuals or entities if we would consider the possibility of asymmetric individual preferences, i.e., asymmetric payouts so that the payout structure depicted in the table above may correspond to different games depending on which player's perspective one takes. There are many biological and social contexts in which asymmetric payouts can occur, however, in this paper we will only consider symmetric settings. 

All the games described above have been used to model cooperation in a wide range of systems, including biological systems, insect societies, animal conflicts and human interactions. If we take, for instance, scenario (a) in figure \ref{fig:scenarios} and focus on the case of an insect colony of cooperators being invaded by a group of insect defectors, the different games described above will represent different possibilities for how much gain both groups can attain depending on the nature of the interactions. For example, in the case that the interactions are modelled using a prisoner's dilemma game then cooperator-defector interactions result in higher payouts for the defectors than in cooperator-cooperator interactions, and vice versa if the interactions are modelled using the stag hunt game. The latter case would be a situation in which invaders are not able to access as many resources (food) from the resident colony as the members of the colony themselves can.

\subsection{Modelling cooperation games with reaction networks}
\label{sec:reactionnetworks}
As advertised in section \ref{sec:introduction}, it is convenient to model games between an arbitrary number $N$ of players using chemical reaction networks, as we now describe. In chemistry, reaction networks are used to formalise the interaction between different species (e.g. types of molecules) in an environment. This environment may contain different species of molecules that combine with each other via certain chemical reactions in order to form other molecules.

A reaction network consists of two types of objects:  species and reactions. The network is given by a pair $(\mathcal{S}, \mathcal{R})$ consisting of a finite set of species $\mathcal{S}= \{n_1, . . ., n_\ell \}$, and a finite set of reactions $\mathcal{R} = \{R_1, \ldots, R_l\}$ with corresponding reaction rates $\mathcal{K}=\{k_1,...,k_l\}$. Each reaction $R_i$ is modelled as a pair $(A, B)$ of multisets of elements of $\mathcal{S}$. Intuitively, $A$ is the species that are used by the reaction $R_i$ and $B$ is the set of species produced by it. $A$ and $B$ are taken to be multisets, as more than one instance of a species might be required for a reaction $R_i$ to take place. A reaction $R_i=(A, B)$ is also denoted by $A \xrightarrow{k_i} B$.

We are interested in studying populations of strategies, i.e., populations of individuals where each individual is of a type determined by the strategy or action they employ. Individuals interact by playing one of the games discussed in section \ref{sec:cooperationgames}. The interactions of strategies in this setting can be formulated in the form of reaction networks \cite{veloz2014reaction}. The network will then include two species corresponding to the two strategy types that we consider. We denote a player using the \emph{defect} strategy by a D and a player using the \emph{cooperate} strategy by a C. The interaction between these species will produce the payouts (for cooperation and defection) as specified by the game. To keep track of the payout obtained by each strategy, we introduce $G_d$, the gain of defecting and $G_c$, the gain of cooperating, which constitute the other two species in our network such that $\mathcal{S}= \{C,D,G_c,G_d\}$. The resulting reaction network corresponding to the cooperation games of section \ref{sec:cooperationgames} for scenario (a) of figure \ref{fig:scenarios} is given by
\begin{align}\label{eq:genprionsersreactions}
    &R_1: C + C \xrightarrow{k_1} C + C + 2r G_c, \\
    &R_2: C + D \xrightarrow{k_2} C + D + \tau G_d + s G_c, \\
    &R_3: D + C \xrightarrow{k_3} D + C + \tau G_d + s G_c, \\
    &R_4: D + D \xrightarrow{k_4} D + D + 2p G_d.
\end{align}
Here the constants $k_1, k_2, k_3, k_4$ are the reaction rates, i.e. how frequent these reactions actually take place. Throughout this paper, we focus on situations in which $k_2=k_3$. Reaction $R_1$, for instance, describes the interaction between two cooperators leading to $2r$ "molecules" of cooperator gain $G_c$, since according to the payout table of section \ref{sec:cooperationgames} each cooperator receives a payout $r$. The same logic applies to the other reactions.

\begin{figure}[h!]
    \centering
    \includegraphics[width=0.5\textwidth]{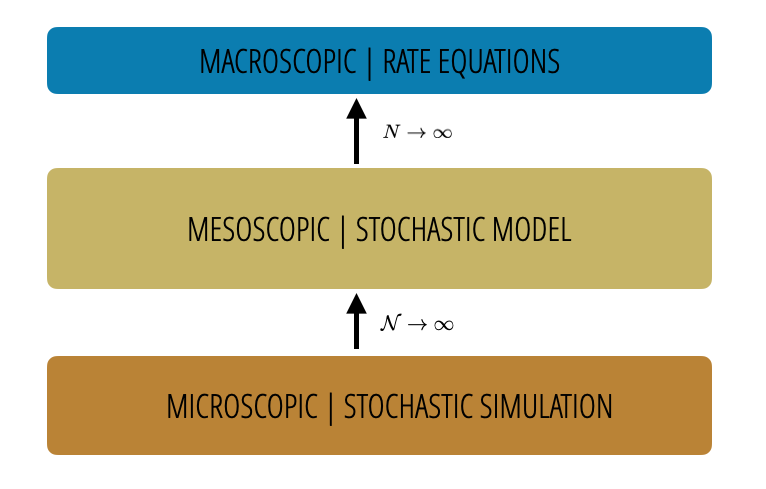}
    \caption{Diagrammatic representation of the different levels of description of the system \eqref{eq:genprionsersreactions} which we employ in this paper.}
    \label{fig:levels}
\end{figure}

As in the context of chemical reactions, the dynamical system described by the reaction network \eqref{eq:genprionsersreactions} is a microscopic stochastic system. In the case of homogeneously mixed populations, which we focus on in this paper, two players are selected randomly from the available pool of $N$ players $(C,D)$ in order to play a game (reaction) in accordance with the reaction rates/frequencies $k_1,k_2,k_3,k_4$ of such games. This random selection of players and reactions introduces (demographic) stochasticity into the system. If we run a number $\mathcal{N}$ of sequences of such games and perform an ensemble average over $\mathcal{N}$ sequences, in the limit $\mathcal{N}\to \infty$ we can describe the system using a coarse-grained (mesoscopic) description in terms of probability distributions (see figure \ref{fig:levels}).
In this mesoscopic level of description we can formulate an analytic stochastic model as will be described in \ref{sec:stochasticmodel}. As usual, the stochastic effects wash away at long time scales ($t\to\infty$) as well as for a very large number of players  (i.e. in the thermodynamic limit $N\to\infty$). In the latter limit, the dynamics of the system reduces to classical rate equations. In section \ref{sec:simulation} we will describe in greater detail how to simulate the reaction network \eqref{eq:genprionsersreactions} at the microscopic level. Below, we formulate and extract useful information from the other two levels of description.

\subsubsection{Macroscopic rate equations}
\label{sec:rateequations}
As mentioned above, for a large number of players $N\to\infty$, the dynamics of the network \eqref{eq:genprionsersreactions} is determined by classical rate equations. As is typical in the context of chemical systems, given the network \eqref{eq:genprionsersreactions} one can readily apply the law of mass action, and assuming $k_2=k_3$, the kinetics result in two uncoupled ordinary differential equations (ODEs) for the net payouts,
\begin{align*}
    \dot{g_c} &= 2\kappa_1r c^2 + 2 \kappa_2 s cd, \\
    \dot{g_d} &= 2 \kappa_2 \tau cd + 2\kappa_4p d^2.
\end{align*}
Here the quantities $g_c, g_d, c$ and $d$ represent the average density of quantities/populations of $G_c, G_d, C, D$ in \eqref{eq:genprionsersreactions} and when $N\to\infty$.  For instance $g_c=\langle \frac{G_c}{N}\rangle|_{N\to\infty}$. Similarly, $\kappa_1,\kappa_2,\kappa_3$ are average reaction rates for the densities $g_c, g_d, c$ and $d$, i.e. $\kappa_i=\langle k_i N\rangle|_{N\to\infty}$. The "dot" appearing in the various quantities stands for the time derivative $\partial/\partial t$, for instance $\dot g_c=\frac{\partial}{\partial t} g_c$. We note that the total number of cooperators $c$ and defectors $d$ is constant in time and hence equal to the microscopic quantities, that is $c=\langle \frac{C}{N}\rangle|_{N\to\infty}=\frac{C}{N}$ and $d=\langle \frac{D}{N}\rangle|_{N\to\infty}=\frac{D}{N}$. The ODEs above are the same as those appearing in \cite{veloz2014reaction} with an appropriate choice of parameters.

The population of species $g_c$ and $g_d$ at time $t$, correspond to the total gain from the cooperation and defection generated in the interaction between the players. The profit generated per cooperative decision is simply $p_c=g_c/c$ and for the defector decision $p_d=g_d/d$, assuming that the gain from cooperation and defection is divided equally between agents that follow the corresponding strategy. It is straightforward to solve these simple ODEs assuming zero gains as an initial condition, yielding
\begin{equation}\label{eq:profit}
    p_c(t) = (2\kappa_1r c + 2\kappa_2 s d) t~~, ~~ p_d(t) = (2\kappa_2 \tau c + 2\kappa_4p d) t. 
\end{equation}
It is useful to compare the profits acquired by cooperators and defectors by computing the profit ratio
\begin{equation}\label{profitratio}
    p_r =\frac{p_c(t)}{p_d(t)}= \frac{\kappa_{1} r c + \kappa_2sd}{\kappa_{4} p d + \kappa_{2} \tau c}.
\end{equation}
If we assume that the principle governing group evolution is the maximisation of the average profits then from $p_r$ we deduce that cooperation is favoured if $p_r >1$ (which is a possible scenario for many choices of parameters) and defection is preferred otherwise. We will return to this analysis in more detail in section \ref{sec:Nash}. The average profits are the only quantities of interest that can be extracted from the macroscopic rate equations, and hence this macroscopic level of description does not contain additional information about the reaction network. Below, we formulate the mesoscopic stochastic model depicted in figure \ref{fig:levels}.

\subsection{Stochastic formulation of cooperation games} 
\label{sec:stochasticmodel}
The mesoscopic stochastic model of the chemical reaction network \eqref{eq:genprionsersreactions} can be formulated using methods from quantum mechanics applied to the notion of stochastic Petri nets for complex systems, as described in \cite{Baez_2017} and which we briefly define. A stochastic Petri net consists of a set of species $\mathcal{S}$ and a set of transitions $\mathcal{R}$ together with the existence of three functions: $f_i:\mathcal{S}\times\mathcal{R}\to\mathbb{N}$ where $\mathbb{N}$ is the set of natural numbers, specifies how many copies of each species appears as input for each transition; $f_o:\mathcal{S}\times\mathcal{R}\to\mathbb{N}$ specifies how many copies of each species appears as output for each transition; and $f_k:\mathcal{R}\to(0,\infty)$ specifies the rate $k_i$ for each transition. From this point of view, it is obvious that  \eqref{eq:genprionsersreactions} can be regarded as a stochastic Petri net.

Associated to each stochastic Petri net is a master equation that specifies how the probability that we have a given number of each species changes with time. In particular, consider a reaction network consisting of $\ell$ species  $\mathcal{S}=\{s_1, \ldots, s_\ell\}$ and reactions $\mathcal{R}=\{R_1, \ldots, R_l\}$ with corresponding rates $\mathcal{K}=\{k_1, \ldots, k_l\}$. A given state of the population at a particular point in time is characterised by the number of individuals of a given a species $\vec{n}=(n_{s_1}, n_{s_2}, \ldots, n_{s_\ell})$ while the generating function
\begin{equation} \label{eq:genfundef}
    P(t, \vec{z}) = \sum_{\vec{n}} p_{\vec{n}} (t) z_{s_1}^{n_{s_1}} z_{s_2}^{n_{s_2}} \ldots z_{s_\ell}^{n_{s_\ell}} \equiv \sum_{\vec{n}} p_{\vec{n}} (t) \vec{z}^{\,\vec{n}} \,,
\end{equation}
gives the stochastic characterisation of the population, where we have defined $\vec{z}^{\,\vec{n}}=z_{s_1}^{n_{s_1}} z_{s_2}^{n_{s_2}} \ldots z_{s_\ell}^{n_{s_\ell}}$. In particular, the generating function $P(t, \vec{z})$ states that at time $t$ the population is at state $\vec{n}$ with probability $p_{\vec{n}}(t)$. Here the variable $z_{s_i}$ is interpreted as a \emph{creation operator} in the quantum language, in particular $a_{s_i}^\dagger = z_{s_i}$ is an operator that adds one individual of the species $s_i$ to the population, and $a_{s_i} = \partial_{s_i} \equiv \frac{\partial}{\partial z_{s_i}}$ is an \emph{annihilation operator} removing one individual of the species $s_i$ from the population. The creation and annihilation operators satisfy the commutation relations
\begin{equation}
[a_{s_i}, a^\dagger_{s_j}] = \delta_{s_i s_j}~,
\end{equation}
where $\delta_{s_is_j}$ the Kronecker delta.

The dynamics of the stochastic system is determined by the dynamics of the generating function $P(t, \vec{z})$ which satisfies the master equation
\begin{equation}\label{eq:mastereqn}
    \frac{d}{d t} P(t, \vec{z}) = H P(t, \vec{z})\,~~,
\end{equation}
akin to the evolution equation for a density matrix in quantum mechanics, where $H$ is the Hamiltonian of the system (or the infinitesimal generator) and constructed from the reaction rates $k_i$ and the creation/annihilation operators $a_{s_i}^\dagger, a_{s_i}$. In particular, if we let $\alpha_{ij}$ (res. $\beta_{ij}$) be the number of individuals of species $i$ that are consumed (res. produced) by reaction $R_j$ on a stochastic Petri net, the Hamiltonian $H$ takes the general form \cite{Baez_2017}
\begin{equation}\label{eq:generalH}
H= \sum_{j=1}^{l} k_j H_{R_j}~~,~~H_{R_j}= \Pi_{i=1}^{\ell} (a_{s_i}^{\dagger})^{\beta_{ij}} \cdot \Pi_{i=1}^{\ell} a_{s_i}^{\alpha_{ij}} - \Pi_{i=1}^{\ell} (a_{s_i}^{\dagger})^{\alpha_{ij}} \cdot \Pi_{i=1}^{\ell} a_{s_i}^{\alpha_{ij}}~~,
\end{equation}
where $H_{R_i}$ is  the contribution to the Hamiltonian for reaction $R_i$. At this point, we have described a general prescription for turning chemical reactions networks into stochastic Petri nets. Below, we introduce the specific stochastic models that we study throughout this paper.

\subsubsection{Mesoscopic stochastic model}
\label{sec:mesomodel}
Given the general considerations on stochastic Petri nets described above, we now turn the specific chemical reaction network \eqref{eq:genprionsersreactions} for scenario (a) in figure \eqref{fig:scenarios} into a stochastic model for cooperation games. As mentioned earlier, stochastic Petri nets require that $f_i, f_o\in \mathbb{N}$ which, for the reaction network \eqref{eq:genprionsersreactions}, implies that the payouts must be non-negative, i.e., $\tau, r, p, s \geq 0$. This can in general be achieved by a rescaling of the payouts. In fact, we are always able to rescale the payouts so that $\min\{\tilde{\tau}, \tilde{r}, \tilde{p}, \tilde{s}\} = 0$ and $\max\{\tilde{\tau}, \tilde{r}, \tilde{p}, \tilde{s}\} = 1$, where $\tilde{\tau}, \tilde{r}, \tilde{p}, \tilde{s}$ are the rescaled payouts. This reduces the number of parameters we have to fix from four to two. However, for clarity we will work with the non-rescaled payouts but impose the condition $\tau, r, p, s \geq 0$, except when otherwise stated.

For the specific reaction network \eqref{eq:genprionsersreactions} we have 4 different species $\mathcal{S}=\{C,D,G_c,G_d\}$ and hence we introduce the total population of each species by $\vec{n} = (n_c, n_d, n_{G_c}, n_{G_d})$. Using these, we can define the generating function \eqref{eq:genfundef} as
\begin{equation} \label{eq:genfundef2}
    P(t, \vec{z}) = \sum_{\vec{n}} p_{\vec{n}} (t) z_c^{n_c}z_d^{n_d}z_{G_c}^{n_{G_c}}z_{G_d}^{n_{G_d}}  \,,
\end{equation}
obeying the master equation \eqref{eq:mastereqn}. In order to completely specify the evolution of the systems, we must provide the Hamiltonian $H$. Using the general prescription of \eqref{eq:generalH} the Hamiltonian for \eqref{eq:genprionsersreactions} takes a simple form
\begin{align}\label{eq:PDgenerator}
    H & =  k_1\left((a^\dagger_{G_c})^{2r} - 1 \right)  (a^\dagger_c)^2 a_c^2 + 2 k_2  \left( (a^\dagger_{G_d})^\tau (a^\dagger_{G_c})^s -1 \right) a^\dagger_c a^\dagger_d a_c a_d   \\
    & \quad  + k_4 \left( (a^\dagger_{G_d})^{2p} - 1  \right)(a^\dagger_d)^2 a_d^2 ~~, \nonumber
\end{align}
where we remind the reader that we have set $k_2=k_3$. Conveniently, the master equation \eqref{eq:mastereqn} in this case can be solved exactly. We consider the initial condition in which we have a $C_0$ number of cooperators and a $D_0$ number of  defectors, as well as vanishing gains. Therefore, at $t=0$ the generating function takes the form $P(t=0, \vec{z}) = z_c^{C_0} z_d^{D_0}$. Since, as we noted in section \ref{sec:reactionnetworks}, the number of cooperators and defectors does not change in time, the solution to \eqref{eq:mastereqn} is simply
\begin{equation}\label{eq:solution}
    P(t, \vec{z}) = e^{H t} P(0, \vec{z}) = e^{\gamma(z_{G_c}, z_{G_d}) t } z_c^{C_0} z_d^{D_0}\,,
\end{equation}
where the function $\gamma(z_{G_c}, z_{G_d})$ is given by
\begin{equation}\label{gamma}
    \gamma(z_{G_c}, z_{G_d}) =  k_1 C_0(C_0-1) \left(z_{G_c}^{2r}  - 1 \right)  + 2k_2 C_0 D_0 \left( z_{G_d}^\tau z_{G_c}^s -1 \right) + k_4 D_0(D_0-1) \left(z_{G_d}^{2p}  - 1 \right) \,.
\end{equation}
From \eqref{eq:solution} and \eqref{gamma} it is obvious that at $\vec{z} = \vec{1} \equiv (1,1,1,1)$, the function $\gamma(z_{G_c}, z_{G_d})$ vanishes and the generating function equals unity. This is equivalent to the sum over the probabilities of all possible population states, as can be seen from \eqref{eq:genfundef}. 

In this stochastic formulation, \eqref{eq:solution} allows us to define and study different observable quantities of the system.  For instance, we can define observables corresponding to the profit rate for each strategy $P_C$ and $P_D$ which count the profit accumulation per unit time, per individual of a given strategy. These observables are the equivalent of the total profits that we defined for the macroscopic rate equations \eqref{eq:profit}. More precisely, in order to compute expectation values of such observables from \eqref{eq:solution}, we define the observable as an operator, act with it on the generating function $P(t, \vec{z})$ and set all $\vec{z}=\vec{1}$ in order to obtain the average profit over all possible states. Therefore, we define the \emph{profit rate operators} $\hat{P}_C$ and $\hat{P}_D$ according to 
\begin{equation}\label{profitoperators}
    \hat{P}_C = \frac{1}{t C_0} N_{G_c} \,, \qquad 
    \hat{P}_D = \frac{1}{t D_0} N_{G_d} \,,
\end{equation}
where $N_{s_i} = a^\dagger_{s_i} a_{s_i} = z_{s_i} \partial_{s_i} $ is the {\em number operator} for species $s_i$ and counts the number of individuals of each species in a given state. Explicitly computing the average profit rates we find
\begin{align} \label{eq:averageprofits}
    \langle \hat{P}_C\rangle & = \left. \hat{P}_C P(t,\vec{z})  \right|_{\vec{z} = \vec{1} } = \left. (\hat{P}_C \gamma) t P(t, \vec{z}) \right|_{\vec{z} = \vec{1} }  = \left( 2 r k_1 (C_0-1) + 2 k_2 s  D_0 \right) \,, \\
    \langle \hat{P}_D \rangle & = \left. \hat{P}_D P(t,\vec{z})  \right|_{\vec{z} = \vec{1} } = \left. (\hat{P}_D \gamma) t P(t, \vec{z}) \right|_{\vec{z} = \vec{1} }  =  \left( 2\tau k_2  C_0 + 2 p k_4 (D_0 - 1 ) \right) \,.
\end{align}
From these average profits, the profit ratio \eqref{profitratio} of the preceding section is recovered in the limit of large number of players. Precisely, we define the fixed fraction of cooperators and defectors according to $c = C_0 /N $ and $d = D_0 / N$ where $N=C_0+D_0$ as well as scale the reaction rates such that $k_i\to\frac{\kappa_i}{N}$ leading to the ratio of profit rates at finite $N$ given by\footnote{The rationale behind this scaling relation for the reaction rates is rooted in analogous contexts in biomolecular reactions in which reaction rates in \eqref{eq:genprionsersreactions} would scale with the volume. In the present context, we wish to guarantee that per unit time we obtain the same number of interactions per individual in the limit $N\to\infty$.} 
\begin{equation} \label{stochprofratio}
    p_r^N =\frac{\langle \hat{P}_C\rangle}{\langle \hat{P}_D \rangle}= \frac{ r \kappa_1 (c N-1) +\kappa_2 s dN}{\kappa_2\tau c N+ p \kappa_4 (d N-1)}\,.
\end{equation}
Given this expression, we can take the thermodynamic limit $N \to \infty$ and find that $p_r^\infty=p_r$ where $p_r$ is given in \eqref{profitratio}. Hence, as advertised, in the thermodynamic limit we obtain the average profit rates resulting from the macroscopic description. 

In addition to the expected values for the profits, it is possible to obtain a lot more detailed information about the profit distribution, which was not accessible using the macroscopic description of section \ref{sec:rateequations}. In particular, we can obtain the variance of the profit distribution by defining appropriate operators, in particular the variance is obtained via
\begin{align}
    {\rm Var}(\hat{P}_C) & = \left. \hat{P}^2_C P(t,\vec{z})\right|_{\vec{z} = \vec{1} }  - \langle \hat{P}_C \rangle^2 = \frac{ \left(4 (C_0-1) k_1 r^2+ 2 D_0 k_2 s^2\right)}{t C_0} \\
    {\rm Var}(\hat{P}_D) & = \left. \hat{P}^2_D P(t,\vec{z})\right|_{\vec{z} = \vec{1} }  - \langle \hat{P}_D \rangle^2 = \frac{ \left(2 C_0 k_2 \tau^2 + 4 (D_0-1) p^2 k_4 \right)}{t D_0} \,,
\end{align}
that is, by acting twice with the profit operators $\hat{P}_X$ with $X \in \{C,D\}$ and subtracting the contribution from the average profit rates. Clearly, the variances scales with time as $1/t$ while the expectation values for the profit rates are constant in time, implying that stochastic effects vanish a very long times $t \to \infty$. In addition, if we perform the same scalings of the various quantities as above \eqref{stochprofratio}, we find a realisation of the central limit theorem ${\rm Var}(\hat{P}_X)^{1/2} \sim 1/\sqrt{N}$ in the thermodynamic limit $N \to \infty$.

Besides the expected values and the variances, it is possible to obtain the moment generating function $M_{P_X}(s_x, t)$ analytically for the profit rate operators $\hat{P}_X$. The moment generating function is defined by the expectation value of the exponential of $\hat P_X$, multiplied by a dual parameter $s_x\in \{s_c,s_d\}$ such that\footnote{The parameter $s_X$ can be interpreted as a dual \emph{chemical potential} to the operator $\hat P_X$.}
\begin{equation}
    M_{P_X}(s_x, t) = \langle e^{s_x \, \hat{P}_X} \rangle\,.
\end{equation}
In appendix \ref{app:LDT}, we show precisely that the moment generating functions for the profit rate operators are given by the general form 
\begin{align}
    M_{P_X}(s_x, t) & =   \exp( \lambda_x(s_x, t) t ) \,,
\end{align}
where the function $\lambda_x(s_x, t)$ reads
\begin{align} \label{eq:momentgena}
    \lambda_c(s_c, t) & =  k_1 C_0(C_0-1) \left(e^{2r\, \frac{s_c}{t C_0}}  - 1 \right)  + 2 k_2 C_0 D_0 \left(  e^{s \, \frac{s_c}{t C_0}} -1 \right)  ~,\\
    \lambda_d(s_d, t) & = 2 k_2 C_0 D_0 \left( e^{\tau\, \frac{s_d}{t D_0}} -1 \right) + k_4 D_0(D_0-1) \left(e^{2p\, \frac{s_d}{t D_0} }  - 1 \right)~~.
\end{align}
Hence, in the context of this simple stochastic model, the moment generating function is exactly calculable and provides insight into all statistical moments of the profit rates. The moment generating function will be crucial in the analysis of the Nash equilibria when considering individuals that are risk-averse, as we will see in section \ref{sec:Nash}. In section \ref{sec:simulation} we will show how the microscopic analysis captures the observables that we defined in this stochastic formulation. Below, we introduce the more general models that can capture scenarios (b) and (c) in figure \ref{fig:scenarios}.

\subsubsection{Stochastic models for more general scenarios}
\label{sec:individualgain}
As mentioned in section \ref{sec:introduction} we are interested in modelling the various scenarios depicted in figure \ref{fig:scenarios}. In the previous sections, we described scenario (a) in detail. However, in order to describe the presence of rogue individuals, it is necessary to formulate a more general chemical reaction network than the one introduced in \eqref{eq:genprionsersreactions}. Since the analysis is very similar to the previous sections, though more cumbersome, we describe these models briefly. The most general model is that of scenario (c) where the gains of each cooperator or defector are not shared among the other individuals but accumulated individually. Therefore, consider the following  reaction network where we distinguish each cooperator $C_i$ and each defector $D_i$ such that 
\begin{align}\label{eq:genprionsersreactionslabeled}
    &R_1: C_i + C_j \xrightarrow{k_1} C_i + C_j + r (G_c^i + G_c^j)~,\forall~ i\ne j~ \\
    &R_2: C_i + D_k \xrightarrow{k_2} C_i + D_k + \tau G_d^k + s G_c^i, \\
    &R_3: D_i + C_k \xrightarrow{k_3} D_i + C_k + \tau G_d^i + s G_c^k, \\
    &R_4: D_k + D_l \xrightarrow{k_4} D_k + D_l + p (G_d^k + G_d^l)~,\forall~ k\ne l~.
\end{align}
In general each of these $C_i$ and $D_i$ can stand for entire groups and $G_c^i$ and $G_d^i$ for the gains associated to each of those groups. Hence, the reaction network above is able to describe games between groups of different sizes. However, we do not consider such possibility here and focus on the case in which each $C_i$ and $D_i$ are composed of just one individual. In this case,  we have that $C_i$ with $i=1, \ldots , C_0$ and $D_k$ with $k=1, \ldots ,D_0$ where $C_0$ and $D_0$ are the total number of cooperators and defectors, respectively. Following the same footsteps as in the case of scenario (a) we find that the Hamiltonian for the reaction network \eqref{eq:genprionsersreactionslabeled} is given by
\begin{align}\label{eq:PDgeneratorlabeled}
    H & =  k_1 \sum_{ \substack{i,j = 1 \\ i \neq j}}^{C_0} \left((a^\dagger_{G_c^i})^{r} (a^\dagger_{G_c^j})^{r} - 1 \right) N_{C_i} N_{C_j} + 2 k_2 \sum_{i=1}^{C_0} \sum_{k=1}^{D_0}  \left( (a^\dagger_{G_d^k})^\tau (a^\dagger_{G_c^i})^s -1 \right) N_{C_i} N_{D_k}   \\
    & \quad  + k_4\sum_{\substack{k,l = 1 \\ i\neq j}}^{D_0} \left( (a^\dagger_{G_d^k})^{p} (a^\dagger_{G_d^l})^{p} - 1  \right) N_{D_k} N_{D_l} \nonumber~~,
\end{align}
where we have again focused on the situation in which $k_2=k_3$ and where we have defined $N_{C_i}=a^\dagger_{C^i}a_{C^i}$ and $N_{D_i}=a^\dagger_{D^i}a_{D^i}$ as the number operators for each type of individuals in the population. The individual creation and annihilation operators now satisfy commutation relations of the form
\begin{equation} \label{comrel}
    [a_{C_i}, a_{C_j}^\dagger] = \delta_{ij} = [a_{G_c^i}, a^\dagger_{G_c^j}]\,,
\end{equation}
and likewise for the operators associated to $D_k$. As we are now tracking each individual, we are interested in the individual profit rate operators defined as 
\begin{equation}
    \hat{P}_{C_i} = \frac{N_{G_c^i}}{t} ~~,~~ \hat{P}_{D_k} = \frac{N_{G_d^k}}{t}~~.
\end{equation}
Similarly to scenario (a) we can also obtain the moment generating function for these profit rate operators analytically.  This can be computed in a similar fashion as outlined in appendix \ref{app:LDT} leading to
\begin{align}
    M_{C_i} (\gamma_i)   = \langle e^{\gamma_i  \hat{P}_{C_i}} \rangle = \exp \left( \lambda_{C_i} (\gamma_i) t \right) ~~,~~M_{D_k} (\delta_k)   = \langle e^{\delta_k  \hat{P}_{D_k}} \rangle = \exp \left( \lambda_{D_k} (\delta_k) t \right) \,,
\end{align}
where the functions $ \lambda_{C_i}(\gamma_i)$ and $\lambda_{D_k}(\delta_k)$ are given by
\begin{align}
    \lambda_{C_i}(\gamma_i) & =  2 k_1 (C_0-1) \left(e^{r\, \frac{\gamma_i}{t}}  - 1 \right)  + 2 k_2 D_0 \left(  e^{s \, \frac{\gamma_i}{t }} -1 \right) \,, \\
    \lambda_{D_k}(\delta_k) & = 2 k_2 C_0  \left( e^{\tau\, \frac{\delta_k}{t}} -1 \right) + 2 k_4 (D_0-1) \left(e^{p\, \frac{\delta_k}{t} }  - 1 \right)\,.
    \label{lambdaDk}
\end{align}
From the general reaction network \eqref{eq:genprionsersreactionslabeled} and the Hamiltonian \eqref{eq:PDgeneratorlabeled} we can recover the reaction network for scenario (a) in \eqref{eq:genprionsersreactions} and the Hamiltonian \eqref{eq:PDgenerator} by simply not distinguishing between cooperators and defectors such that $C_i=C$ and $D_i=D$ for all $i$ and similarly for the associated gains. By the same token we can obtain from the reaction network \eqref{eq:genprionsersreactionslabeled} the reaction network for scenario (b) in which only cooperators $C$ share their profit equally and defectors $D_k$ keep their gains to themselves. This is obtained by not distinguishing between cooperators such that $C_i=C~,~\forall~i$ and similarly for the gains $G_i$. In this case the chemical reaction network is
\begin{align}\label{eq:genprionsersreactionslabtheds}
    &R_1: C + C \xrightarrow{k_1} C + C + 2 r G_c ~,\forall~ i\ne j~ \\
    &R_2: C + D_k \xrightarrow{k_2} C + D_k + \tau G_d^k + s G_c, \\
    &R_3: D_k + C \xrightarrow{k_3} D_k + C + \tau G_d^k + s G_c, \\
    &R_4: D_k + D_l \xrightarrow{k_4} D_k + D_l + p (G_d^k + G_d^l) ~,\forall~ k\ne l~.
\end{align}
In this scenario, the moment generating functions for the profit rate operators $\hat{P}_C$ and $\hat{P}_{D_k}$ take the form
\begin{equation}
    M_C(s_c) = \exp \left( \lambda_c(s_c) t \right) \,, \qquad M_{D_k}(\delta_k) = \exp \left( \lambda_{D_k} (\delta_k) t \right) \,,
\end{equation}
where $\lambda_C(s_c)$ is given in \eqref{eq:momentgena} and $\lambda_{D_k}(\delta_k)$ given in \eqref{lambdaDk} with $k_2=k_3$. This concludes our formulation of the stochastic models. We will now use these results to study Nash equilibria and the emergence of cooperation.

\section{Nash equilibria and the emergence of cooperation}
\label{sec:Nash}
Our interest in this paper, as highlighted in section \ref{sec:introduction}, is to uncover a general mechanism that can lead to cooperative behaviour in sizeable populations. To understand whether or not cooperation occurs in the models that we described in the previous section, we need to understand the Nash equilibria in the dynamics of the reaction networks. However, the very definition of Nash equilibria requires that we prescribe the utility function that every individual or group of individuals wishes to maximise. In this section, we will analyse Nash equilibria using two types of utility functions. The first is based on maximisation of average profits in the long run, which is the typical measure assumed in most literature. The second measure is the novel utility function that we propose, which postulates that individuals are risk-averse and hence wish to maximise their minimum payout with a certain confidence level. We will show that this second measure can lead to cooperative behaviour.

\subsection{Utility based on average profit}
In the models we described in section \ref{sec:basicmodandmachine} the number of cooperators and defectors are fixed in time. Nevertheless, we can obtain the possible Nash equilibria by considering whether there is an incentive for an individual to switch strategies given a set of payout parameters and a specific population composition. To this end, we observe that the expected profits per time for each type of player obtained in \eqref{eq:averageprofits} for scenario (a) can be written as
\begin{align}
\label{eq:profitpertime}
    P_{C}(c):= \langle \hat{P}_C\rangle & =
    \begin{cases}
    2 r \frac{(c N-1)}{N} +  2 s  (1-c) \quad &\text{when}\; c\in [1/N, 1]\\
    0 &\text{else}
    \end{cases}\\
    P_{D}(c):= \langle \hat{P}_D \rangle & =
    \begin{cases}
      2\tau c + 2 p \frac{((1-c)N - 1 )}{N}\quad &\text{when}\; c\in [0, (N-1)/N]\\
    0 &\text{else}~,
    \end{cases}
\end{align}
where $c\in \{n/N \, \vert\,n=0, \ldots, N\}$ is the fraction of cooperating players in the population, so $C_{0}=cN$, and we have set the rates $k_{1}=k_2=k_{3}=k_4=1/N$. The reason for focusing on the case in which all reaction    rates scale as $1/N$ is to not introduce any biases in the model and make it equally likely for every individual to play a game with any other individual. Working with different reaction rates may be interesting but requires an explanation of why, for specific systems, it is more likely, for instance, for cooperators to meet defectors than for defectors to meet other defectors. Such contexts would be equivalent to postulating the pre-existence of selection biases into the population structure. However, we are interested in understanding whether cooperation can occur without biasing the system or postulating effects due to other mechanisms. Given this and the profit rates \eqref{eq:profitpertime} we define Nash equilibrium according to
\begin{definition}[Nash equilibrium] Nash equilibrium is given by a point $c_{n}$ with
\begin{equation}\label{eq:Nash}
    P_{D}(c_{n}) \geq P_{C}(c_{n+1})\quad \text{and}\quad P_{C}(c_{n}) \geq P_{D}(c_{n-1}),
\end{equation}
where $c_{n}:=n/N$.
\end{definition}
This definition expresses that the fact that Nash equilibrium requires that there is no incentive for either cooperators or defectors to change strategies and begin defecting or cooperating, respectively. Before using this definition in order to study specific examples, it is useful to consider a few properties of the average profits. In particular, the expressions in \eqref{eq:profitpertime} are linear functions in $c$ with $\partial/\partial c P_{C}(c) = 2(r - s)$ and $\partial/\partial c P_{D}(c) = 2(\tau - p)$. The profit per time is thus increasing in $c$ for both cooperators and defectors in all the games we consider here. We also have the useful formulae 
\begin{equation}
\begin{split}
    P_{C}\Big(\frac{1}{N}\Big)  = \frac{2s(N-1)}{N}~~,~~P_{D}\Big(\frac{1}{N}\Big)  = \frac{2(p(N-2) +\tau)}{N}~,\\
    P_{C}\Big(\frac{N-1}{N}\Big)  = \frac{2(r(N-2)+s)}{N}~~,~~ P_{D}\Big(\frac{N-1}{N}\Big) = \frac{2\tau(N-1)}{N}~~.
\end{split}
\end{equation}
By using these expressions, we can analyse the Nash equilibria that arise in the three different types of games that we introduced in section \ref{sec:cooperationgames}. We focus mostly on scenario (a) but we comment appropriately about differences with respect to the other two scenarios. 

\subsubsection{Prisoner's dilemma}
As we mentioned at the beginning of section \ref{sec:mesomodel} it is possible to rescale the payouts and to work, without loss of generality, with rescaled payouts. In practice, we can set $\tau=1$ and $s=0$ such that $r>p$. As it is well known, in the single shot prisoner's dilemma game the Nash equilibrium based on individual profit maximisation leads to no cooperative behaviour, that is $c=0$. Here, though we work with group fitness maximisation, we also obtain the following lemma:
\begin{lemma}[Nash equilibrium for Prisoner's dilemma]
In the PD game there is a unique Nash equilibrium at $c=0$.
\end{lemma}
\begin{figure}[h!]
    \centering
    \includegraphics[scale=2.0]{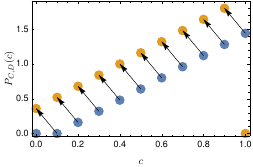}
    \caption{The profit rate of the PD game for defectors (yellow) and cooperators (blue) in a population of $N=10$ as a function of the population composition $c$ with $\tau = 1, r=0.8, p=0.2$ and $s=0$. The arrows indicate that an individual of the type the arrow points from has an incentive to switch to the type the arrow points to.}
    \label{fig:NEPrisonersDilemma}
\end{figure}
\begin{proof}
For $c=0$ to be a Nash Equilibrium of the PD, we must have that
\begin{equation}
    P_{D}(0)=\frac{2(N-1)p}{N} \geq 0 = P_{C}(1/N),
\end{equation}
which is always satisfied since $p>0$. In turn, for $c=1$ to be a Nash equilibrium of the PD, we must have that
\begin{equation}
    P_{D}\Big(\frac{N-1}{N}\Big)=\frac{2(N-1)}{N} \leq \frac{2(N-1)r}{N} = P_{C}(1),
\end{equation}
which is never satisfied since $1>r$. On the other hand, for $c_{n}$ with $1<n<N-1$ to be a Nash equilibrium we must have 
\begin{equation}
\label{eq:NEpd1}
    P_{D}(n/N) = \frac{2(n + (N-n-1)p)}{N}\geq \frac{2nr}{N} = P_{C}((n+1)/N)
\end{equation}
and 
\begin{equation}
\label{eq:NEpd2}
    P_{C}(n/N) = \frac{2(n-1)r}{N}\geq \frac{2(n-1 + (N-n)p)}{N} = P_{D}((n-1)/N).
\end{equation}
However, equations \eqref{eq:NEpd1} and \eqref{eq:NEpd2} cannot be satisfied simultaneously. This completes the proof.
\end{proof}
We illustrate the Nash equilibrium occurring in the prisoner's dilemma in Figure \ref{fig:NEPrisonersDilemma}. As it is clear from the figure, for any non-zero fraction of cooperators $c$, there is always an incentive for cooperators (blue circles) to change their strategy and begin defecting (orange circles). The only possible Nash equilibrium is when $c=0$ and the entire population is composed of defectors. This analysis does not change if we consider scenarios (b) and (c) instead of scenario (a).

\subsubsection{Stag hunt}
\label{sec:SHutility1}
In the SH game, it is again useful to work with rescaled payouts. In particular, we are free to set $r=1$ and $s=0$, without loss of generality, leading to the inequality $\tau>p$. In the two-player stag hunt game, it is known that there are two possible Nash equilibria, one in which both players defect and another in which both players cooperate. We recover this result at the level of two groups of arbitrary players, specifically this leads to the following lemma:
\begin{lemma}[Nash equilibrium for the stag hunt game]
In the SH game there are exactly two Nash equilibria: one at $c=0$ and one at $c=1$.
\end{lemma}
\begin{figure}[h!]
    \centering
    \includegraphics[scale=1.5]{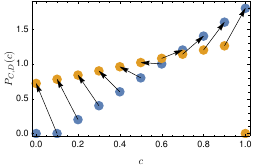}
    \includegraphics[scale=1.5]{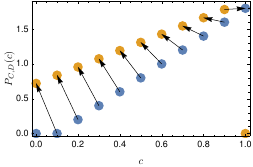}
    \caption{The profit rate of the SH game for defectors (yellow) and cooperators (blue) in a population of $N=10$ as a function of the population composition $c$ with $\tau=0.7, r=1, p=0.4$ and $s=0$ (left) and $\tau=0.99, r=1, p=0.4$ and $s=0$ (right). The arrows indicate that an individual of the type the arrow points from has an incentive to switch to the type the arrow points to.}
    \label{fig:NEStagHunt}
\end{figure}
\begin{proof}
For $c=0$ to be a Nash Equilibrium of the SH game, we must have that
\begin{equation}
    P_{D}(0)=\frac{2(N-1)p}{N} \geq 0 = P_{C}(1/N),
\end{equation}
which is always satisfied since $p>0$. For $c=1$ to be a Nash equilibrium of the SH game, we must have that
\begin{equation}
    P_{D}\Big(\frac{N-1}{N}\Big)=\frac{2(N-1)\tau}{N} \leq \frac{2(N-1)}{N} = P_{C}(1),
\end{equation}
which is always satisfied since $1>\tau$.  On the other hand, for $c_{n}$ with $1<n<N-1$ to be a Nash equilibrium of the SH game we must have 
\begin{equation}
\label{eq:NEsh1}
    P_{D}(n/N) = \frac{2(n\tau + (N-n-1)p)}{N}\geq \frac{2n}{N} = P_{C}((n+1)/N)
\end{equation}
and 
\begin{equation}
\label{eq:NEsh2}
    P_{C}(n/N) = \frac{2(n-1)}{N}\geq \frac{2( (n-1)\tau + (N-n)p)}{N} = P_{D}((n-1)/N).
\end{equation}
However, equations \eqref{eq:NEsh1} and \eqref{eq:NEsh2} cannot be satisfied simultaneously. This completes the proof.
\end{proof}
We illustrate the Nash equilibria occurring in the stag hunt game in figure \ref{fig:NEStagHunt}. As for the PD game, the results presented here also hold for scenarios (b) and (c). We highlight that in contrast with the PD game and the HD game that we analyse below, in the SH game, cooperation is possible when the utility function is the average profits. This shows that while such utility may lead to cooperation in specific games, it does not in general.

\subsubsection{Hawk-dove}\label{sec:HDexpval}
In the HD game we also work with rescaled payouts, namely $\tau=1$ and $p=0$, without loss of generality, leading to the inequality $r>s$. In the two-player HD game, there are at most two Nash equilibria in which one player is cooperating, and the other is defecting. Indeed, we recover this result at the group level, in particular we have the following lemma:
\begin{lemma}[Nash equilibrium for the Hawk Dove game]
In the HD game there is exactly one Nash equilibria at 
\begin{equation}
    n = \Big\lfloor \frac{Ns - r + 1}{1-r+s}\Big\rfloor \quad \text{if}\quad \frac{Ns - r + 1}{1-r+s} \notin \mathbb{Z}
\end{equation}
and there are exactly two Nash equilibria at 
\begin{equation}\label{eq:2Nash}
    n_{1} =  \frac{Ns - s}{1-r+s} \quad \text{and}\quad  n_{2} =  \frac{Ns - r + 1}{1-r+s}\quad \text{if}\quad \frac{Ns - r + 1}{1-r+s} \in \mathbb{Z}.
\end{equation}
\end{lemma}
\begin{figure}[h!]
    \centering
    \includegraphics[scale=1.5]{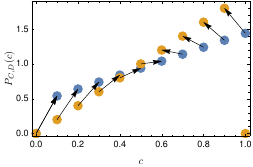}
    \includegraphics[scale=1.5]{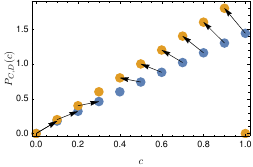}
    \caption{The profit rate of the HD game for defectors (yellow) and cooperators (blue) in a population of $n=10$ as a function of the population composition $c$ with $\tau=1, r=0.8, p=0$ and $s=0.3$ (left) and $\tau=1, r=0.8, p=0$ and $s=0.1$ (right). The arrows indicate that an individual of the type the arrow points from has an incentive to switch to the type the arrow points to.}
    \label{fig:NEHawkDove}
\end{figure}
\begin{proof}
For $c=0$ to be a Nash Equilibrium of the HD game, we must have that
\begin{equation}
    P_{D}(0)=\frac{2(N-1)p}{N} \geq \frac{2(N-1)s}{N} = P_{C}(1/N),
\end{equation}
which is never satisfied since $p<s$. For $c=1$ to be a Nash equilibrium of the HD game, we must have that
\begin{equation}
    P_{D}\Big(\frac{N-1}{N}\Big)=\frac{2(N-1)}{N} \leq \frac{2(N-1)r}{N} = P_{C}(1),
\end{equation}
which is never satisfied since $r<1$.  For $c_{n}$ with $1<n<N-1$ to be a Nash equilibrium of the HD game, we must have 
\begin{equation}
\label{eq:NEhd1}
    P_{D}(n/N) = \frac{2n}{N}\geq \frac{2(n(r-s)+(N-1)s)}{N} = P_{C}((n+1)/N)
\end{equation}
and 
\begin{equation}
\label{eq:NEhd2}
    P_{C}(n/N) = \frac{2((n-1)r +(N-n)s)}{N}\geq \frac{2(n-1)}{N} = P_{D}((n-1)/N).
\end{equation}
Equations \eqref{eq:NEhd1} and \eqref{eq:NEhd2} are satisfied when 
\begin{equation}
\label{eq:NEhd3}
    \frac{Ns - s}{1-r+s} \leq n \leq \frac{Ns - r + 1}{1-r+s}. 
\end{equation}
The difference between the upper and lower bounds of equation \eqref{eq:NEhd3} is one. We also have that 
\begin{equation}
    1<\frac{Ns - r + 1}{1-r+s}<N \quad\text{and}\quad 0<\frac{Ns - s}{1-r+s}<N,
\end{equation}
so that there always exits at least one integer $n$ satisfying the condition. 
\end{proof}
In figure \ref{fig:NEHawkDove} we depict the Nash equilibria in the HD game, which always involve a non-zero fraction of cooperators and defectors. The same results hold for scenarios (b) and (c) in figure \ref{fig:scenarios}. Note that the possibility of two adjacent Nash equilibria here is a finite size effect. As we take the population size to infinity, these equilibria will merge. Note also that this does not contradict the fact that there is a single Nash equilibrium in the two player case ($N=2$), as the condition for the existence of a second Nash equilibrium \eqref{eq:2Nash} requires the population size to satisfy the inequality $N>2$. Of the different games we study here, the HD game is the only one in which, assuming a utility function based on profit maximisation, finite size effects change the Nash equilibria structure.

\subsection{Utility based on risk aversion}
\label{sec:riskaversion}
In the previous section, we assumed that the utility of an individual in our population was given by the expected profit. The actually realised profit, however, is a random variable. If our population consists of individuals that are risk-averse, they may seek to maximise another utility function in a given time range. A risk-averse utility function can be defined as the minimum payout $p_X^{\rm min}$ they would receive with a given level of confidence $\alpha$. This leads to the definition
\begin{definition}[Risk-averse utility]
\label{def:RAutilityoriginal}
The risk-averse utility with risk aversion parameter $\alpha$ is the payout $p_X^{\rm min}$, for which the probability of receiving a payout greater or equal to $p_X^{\rm min}$ is $\alpha$, i.e., \begin{equation}
    \mathbb{P}[P_X(c) \geq p_X^{\rm min}] = \alpha.
\end{equation}
\end{definition} 
This gives us a generalized notion of risk aversion; a risk-averse agent is defined as one that prefers a sure bet over an uncertain one that has a higher expected utility but a minimum payout lower than what is guaranteed by the sure bet. Definition \ref{def:RAutilityoriginal} generalizes this to an agent who prefers an ``almost'' sure bet to one that has a higher expected utility but a lower ``almost'' guaranteed payout. The threshold at which the agent would consider a payout as ``almost'' certain ($\alpha$ in Definition \ref{def:RAutilityoriginal}) determines the agent's degree of risk aversion. The highest level of risk aversion is realized when $\alpha=1$, which corresponds to the case in which the players' utility is given by the maximin value. Here, the agent chooses the action that gives the highest 100\%-guaranteed payout. With a threshold smaller than 1, say $\alpha=0.8$, the agent will choose the action that gives the highest 80\%-guaranteed payout.

The difficult task at hand is to explicitly compute $p_X^{\rm min}$ for a given distribution of payouts and a given fraction $c$ of cooperators. It is possible to have some analytic control by parameterizing the risk aversion based on the concept of Entropic Value-at-Risk introduced in \cite{ahmadi2012entropic}. In particular, it was noted in \cite{ahmadi2012entropic} that the probability that the payout is less than a value $p_X^{\rm min}$ is bounded using Chernoff's inequality, 
\begin{equation}
\mathbb{P}[P_{X}(c) \leq p_X^{\rm min}] \leq \eee^{-s_{x} p_X^{\rm min}} M_{P_{X}(c)}(s_{x}) \quad \text{for all } \quad s_{x}<0~,
\end{equation}
where $ M_{P_{X}(c)}(s_{x})$ is the moment generating function of $P_{X}(c)$. We see precisely the appearance of the moment generating function which, for the models we study here, we computed in section \ref{sec:stochasticmodel}. We can solve for $p_X^{\rm min}$ in the inequality $\eee^{-s_{x}p_X^{\rm min}}M_{P_{X}(c)}(z) \geq 1-\alpha$, yielding a lower bound on $p_X^{\rm min}$ such that
\begin{equation}
    p_X^{\rm min} \geq s_{x}^{-1}\ln\left[(M_{P_{X}(c)}(s_{x})/(1 - \alpha) \right]~.
\end{equation}
The utility is hence lower bounded by the maximal value of the right-hand side in the above expression. As the inequality holds for all values of $s_x <0$, the most stringent lower bound is obtained by the supremum over $s_x<0$. This gives a conservative estimate of the utility, which we define as the conservative risk-averse utility according to
\begin{definition}[Conservative risk-averse utility]
\label{def:RAutility}
The utility of an agent of type $X\in\{C, D\}$ with risk-averseness $\alpha$ in a population of concentration $c$ is lower bounded by the conservative risk-averse utility $U^\alpha_X(c)$:
\begin{equation}
    U^{\alpha}_{X}(c) = \sup_{s_{x}<0}\{p_X^{\rm min}(1-\alpha, s_{x})\}.
\end{equation}
with
\begin{equation}
    p_X^{\rm min}(1-\alpha, s_{x}) = s_{x}^{-1}\ln \left[ M_{P_{X}(c)}(s_{x})/(1-\alpha)\right]~.
\end{equation}
\end{definition}
On the other hand, we can also find an upper bound for the utility $p_X^{\rm min}$ for which $\mathbb{P}[P_X(c) \geq p_X^{\rm min}] = \alpha$ by using the Chernoff bound in the other direction, specifically
\begin{equation}
\mathbb{P}[P_{X}(c) \geq p_X^{\rm min}] \leq \eee^{-s_{x} p_X^{\rm min}} M_{P_{X}(c)}(s_{x}) \quad \text{for all } \quad s_{x}>0\,.
\end{equation}
By using steps similar to the ones above, one can show that $p_X^{\rm min}$ is bounded as
\begin{equation}\label{abounds}
    U^{\alpha}_{X}(c) \leq p_X^{\rm min} \leq V^{\alpha}_{X}(c)~~,
\end{equation}
where 
\begin{equation}
     V^{\alpha}_{X}(c) =  \inf_{s_{x}>0}\left\{ \frac{1}{s_x} \ln \left( \frac{M_{P_X(c)}(s_x)}{\alpha}  \right) \right\} \,.
\end{equation}
One of the important aspects of the risk utility function is that it is a time dependent function. This is clear by looking at the form of the moment generating function $M_{P_X(c)}$ appearing in \eqref{eq:momentgena}. This means that we must specify a timescale on which the players wish to maximise their utility. As advertised in section \ref{sec:introduction} this makes sense in biological contexts, for instance, in which living systems may wish to guarantee a minimum of resources during a specific stage of development. It is worth noting that at very long time scales ($t \to \infty$), both bounds of the utility function in \eqref{abounds} become the expected profit rates \eqref{eq:profitpertime} used in the previous section. This means that the risk utility function defined here is continuously connected to the utility function based on average profits. However, if the individuals are interested in maximising their payouts on a shorter time interval $\Delta t$, random fluctuations will influence their actually realised profit. The conservative utility function \eqref{eq:RAutilitymodel} with $t= \Delta t$ then gives the best possible bound on the minimum value of the profit rate an individual may expect to receive with confidence $\alpha$. For the model we are considering, in particular for scenario (a), we have
\begin{align}
\label{eq:RAutilitymodel}
    U^{\alpha}_{X}(c) &= \sup_{s_{x}<0}\{s_{x}^{-1}( \lambda(s_{x}, \Delta t) \Delta t - \ln(1-\alpha))\}.
\end{align}

Note that the Conservative risk-averse utility of definition \ref{def:RAutility} satisfies the usual conditions for a risk measure to be considered coherent. In particular, it is translation invariant, subadditive, monotonic and positive homogeneous (see \cite{ahmadi2012entropic}). This is in contrast to other measures of risk, such as Value-at-risk (which is not subadditive) and mean-standard-deviation (which is not monotonic). Other coherent risk measures are the worst-case and the Conditional Value-at-risk introduced in \cite{rockafellar2000optimization, rockafellar2002conditional}. Our motivation for using the Entropic Value-at-risk is its explicit use of the moment generating function.

In the remainder of this section, we will analyse the Nash equilibrium structure of the various population games using the conservative risk-averse utility function \eqref{eq:RAutilitymodel}. We note that Nash equilibrium is now defined as in \eqref{eq:Nash} but with $P_X(c)$ exchanged by $p_X^{\rm min}(c)$ for a given confidence level $\alpha$. We also note that we cannot calculate the supremum in equation \ref{eq:RAutilitymodel} analytically, but we can calculate it numerically given all model parameters. In section \ref{sec:simulation} we will compute $p_X^{\rm min}$ numerically instead of working with lower bounds on $p_X^{\rm min}$. Below, we also note the various differences for scenarios (b) and (c).

\subsubsection{Prisoner's dilemma}
\begin{figure}[h!]
    \centering
    \includegraphics[scale=0.48]{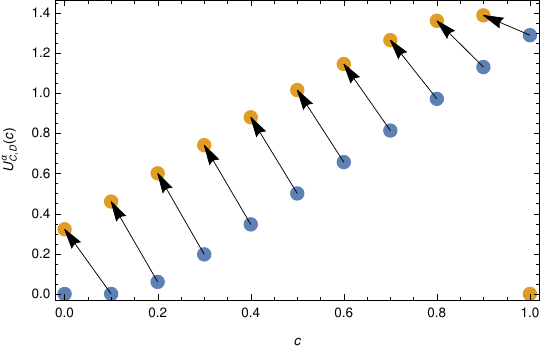}
    \includegraphics[scale=0.48]{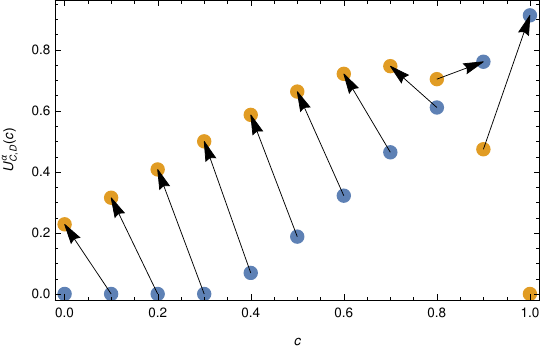}
    \includegraphics[scale=0.48]{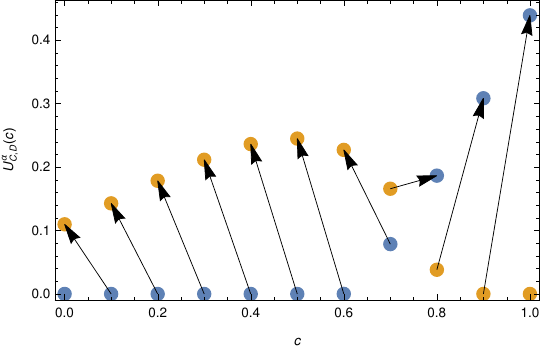}
    \caption{The utility for the PD game for scenario (a) based on Definition \ref{def:RAutility} (with $\Delta t = 1$) for defectors (orange) and cooperators (blue) in a population of $N=10$ as a function of the population composition $c$ with $\tau=1, r=0.8, p=0.2, s=0$ and $\alpha=0.05, 0.5, 0.95$ (left, centre and right respectively). The arrows indicate that an individual of the type the arrow points from has an incentive to switch to the type the arrow points to.}
    \label{fig:NEprisonersdilemmaRA}
\end{figure}

In figure \ref{fig:NEprisonersdilemmaRA} we see that the risk-averse utility leads to the possibility of a second Nash equilibrium in the Prisoner's dilemma at $c=1$ for a timescale of $\Delta t=1$ and for scenario (a). 
At first this result seems counter-intuitive as one would expect a defector to thrive in a community where they are the only defector. The existence of the Nash equilibrium is due to the large variance a single defector faces when they are the only defector (or when there are very few defectors), which comes from the possibility that the defector is not chosen to play. If a cooperator is not chosen to play in such a population, they may still obtain profit from the gain by all the other cooperators. The defector has no such ``safety net'' to rely on. From this, and as earlier advertised, we can conclude that risk aversion can lead to cooperation. 
\begin{figure}[h!]
    \centering
    \includegraphics[scale=0.8]{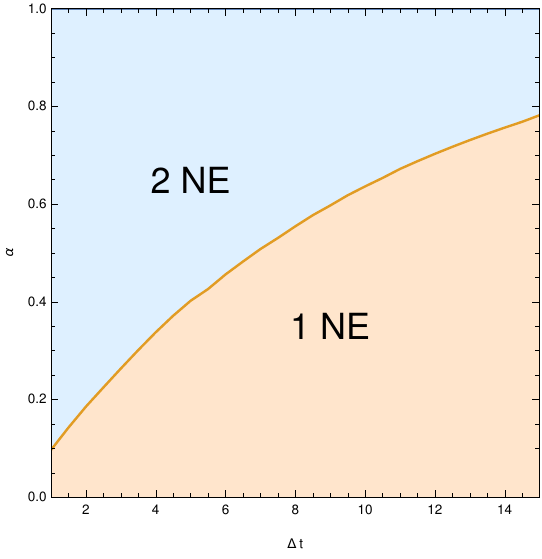}
    \caption{A `phase diagram' for the PD game with $N=10, \tau=1, r=0.8, p=0.2, s=0$. The orange region corresponds to values of $\alpha$ and $\Delta t$ for which only one Nash equilibrium (NE) at $c=0$ (the all $D$ population) exists. In the blue region, a second Nash equilibrium at $c=1$ (the all $C$ population) emerges.}
    \label{fig:PDphasediagram}
\end{figure}
Figure \ref{fig:PDphasediagram} shows for which values of $\alpha$ and $\Delta t$ the new Nash equilibrium at $c=1$ (corresponding to a totally cooperating population) emerges. As expected, the region shrinks when the timescale on which players base their utility $\Delta t$ is increased. However, the new Nash equilibrium is present for a large range of $\alpha$ and $\Delta t$ values, showing that it is a robust feature of the model. We also note that the phase diagram in \ref{fig:PDphasediagram} also indicates that for a given $\Delta t$, there is a corresponding threshold for $\alpha$ (the phase separation curve) above which cooperation emerges. This threshold can presumably dependent on the context and could, for instance, be related to the minimum resources needed for a given insect in a colony.
\begin{figure}
    \centering
    \includegraphics[scale=0.48]{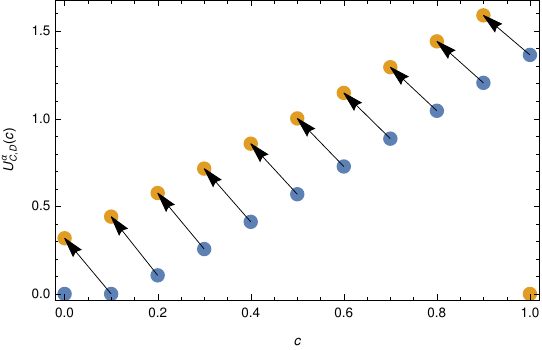}
    \includegraphics[scale=0.48]{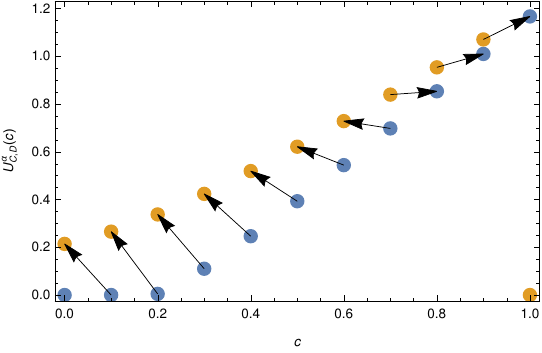}
    \includegraphics[scale=0.48]{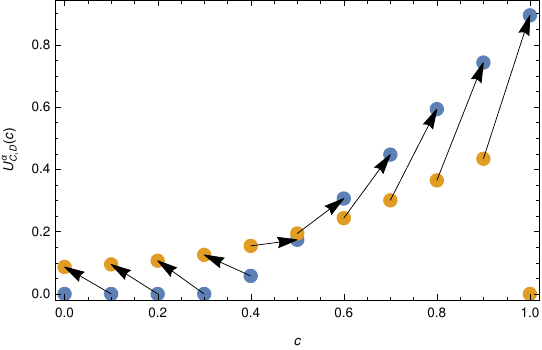}
    \caption{The risk-averse utility for the PD game for scenario (b) (with $\Delta t = 4$) for defectors (orange) and cooperators (blue) in a population of $N=10$ as a function of the population composition $c$ with $\tau=1, r=0.8, p=0.2, s=0$ and $\alpha=0.05, 0.5, 0.9$ (left, centre and right respectively).}
    \label{fig:NEprisonersdilemmaRAalpha}
\end{figure}
In turn, if we look at scenario (b) in which we consider a group of cooperators playing against rogue individuals, the results depicted in figure \ref{fig:NEprisonersdilemmaRAalpha} are qualitatively the same as those of figure \ref{fig:NEprisonersdilemmaRA} for just two groups of cooperators and defectors and the phase diagram is identical to \ref{fig:PDphasediagram}. However, in scenario (b) the effects of risk aversion are more pronounced since the emergence of the second Nash equilibria appears for much lower fractions $c$ of cooperators compared with scenario (a). We note that in both scenarios (a) and (b), the emergence of the second Nash equilibrium appears for any population size with $N\ge3$.

When the utility is based on the model \eqref{eq:genprionsersreactionslabeled} for scenario (c) where the cooperators and defectors do not share their gains with other players, but instead keep their gain individually, the new Nash equilibrium is not present. This implies that the second Nash equilibrium is a consequence of group fitness (i.e. sharing of profits among individuals of like species) combined with risk aversion. 

\subsubsection{Stag hunt}
\begin{figure}[h!]
    \centering
    \includegraphics[scale=0.48]{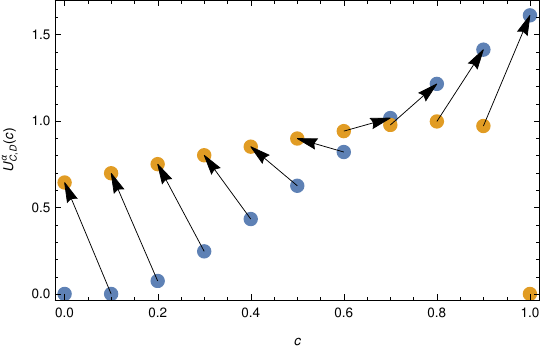}
    \includegraphics[scale=0.48]{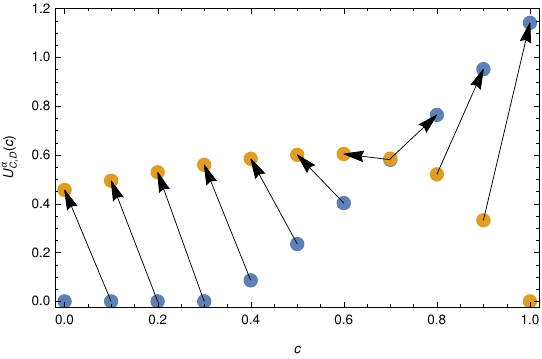}
    \includegraphics[scale=0.48]{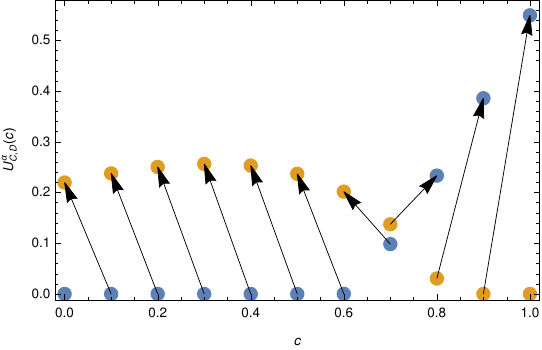}
    \caption{The utility for the SH game based on Definition \ref{def:RAutility} for scenario (a) with $\Delta t = 1$ for defectors (orange) and cooperators (blue) in a population of $n=10$ as a function of the population composition $c$ with $\tau=0.7, r=1, p=0.4, s=0$ and $\alpha=0.05, 0.5, 0.95$ (left, centre and right respectively).}
    \label{fig:NEstaghuntRAalpha}
\end{figure}

In figure \ref{fig:NEstaghuntRAalpha} we see the effect that risk aversion as defined in Definition \ref{def:RAutility} has on the Nash equilibrium structure of the SH game for scenario (a). In this case, the addition of risk aversion does change the shape of the utility functions, but it does not have a significant effect on the Nash equilibrium analysis. We note that in scenarios (b) and (c) the results are qualitatively similar to figure \ref{fig:NEstaghuntRAalpha}.
It is important to note that though this new risk-averse measure does not improve the emergence of cooperation in the SH game, it has the interesting property that it also leads to the emergence of cooperation in the same way that the utility based on average profits in section \ref{sec:SHutility1} did. Therefore, the risk aversion utility of Definition \ref{def:RAutility} is generically leading to cooperation.

\subsubsection{Hawk-dove}
\begin{figure}[h!]
    \centering
    \includegraphics[scale=0.48]{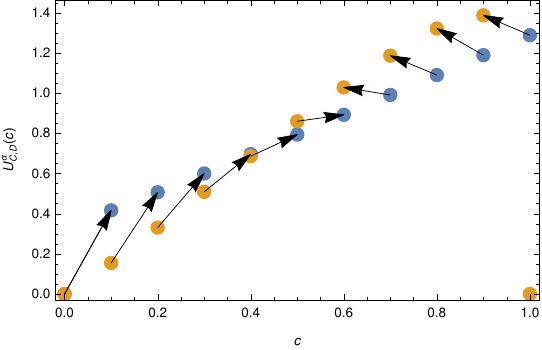}
    \includegraphics[scale=0.48]{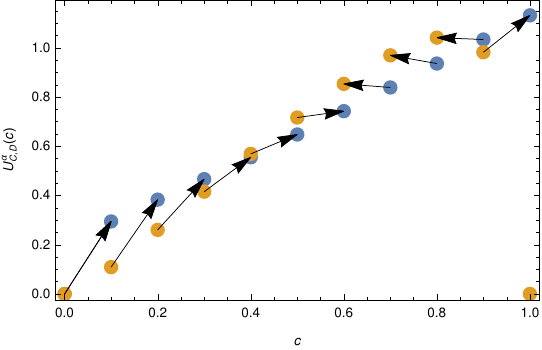}
    \includegraphics[scale=0.48]{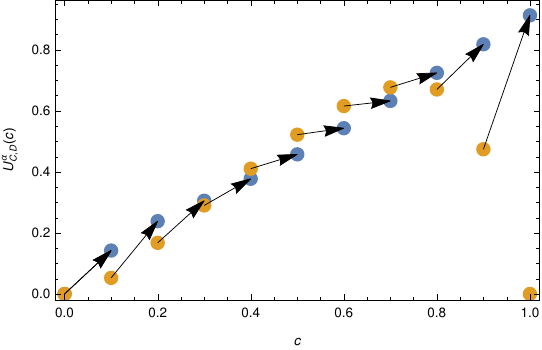}
    \caption{The utility for the HD game based on Definition \ref{def:RAutility} for scenario (a) with $\Delta t =1$ for defectors (yellow) and cooperators (blue) in a population of $N=10$ as a function of the population composition $c$ with $\tau=1.0, r=0.8, p=0, s=0.3$ and $\alpha=0.05, 0.2, 0.5$ (left, centre and right respectively).}
    \label{fig:NEhawkdoveRAalpha}
\end{figure}

\begin{figure}[h!]
    \centering
    \includegraphics[scale=0.725]{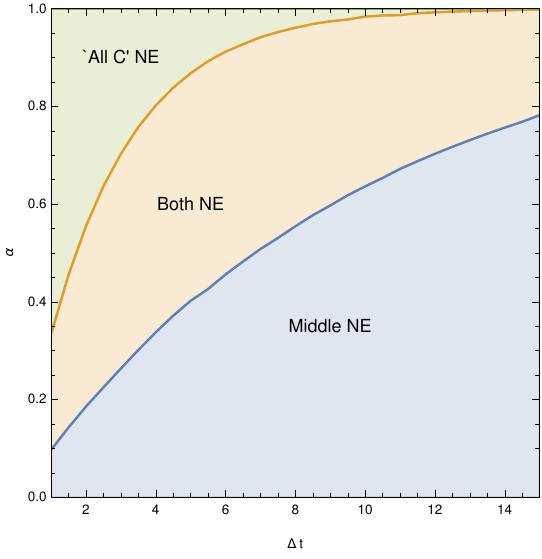}
    \includegraphics[scale=0.725]{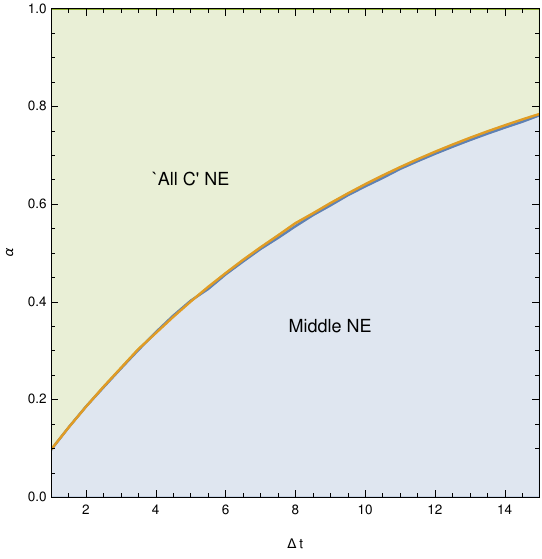}
    \caption{Phase diagrams for the HD game with $n=10, \tau=1, r=0.8, p=0, s=0.3$.  The green region corresponds to values of $\alpha$ and $\Delta t$ with a single Nash equilibrium at the all $C$ population ($c=1$). In the orange region, two NE exist, one at the all $C$ population and a second NE for a non-zero fraction of cooperators in the population. In the blue region, only the Nash equilibrium the non-zero fraction of cooperators exists. The left panel corresponds scenario (a), whereas the right panel corresponds scenario (b).}
    \label{fig:HDphasediagram}
\end{figure}

For the HD game, we see in figure \ref{fig:NEhawkdoveRAalpha} that the inclusion of risk aversion has a strong effect on the Nash equilibrium structure for scenario (a). A small value of $\alpha$ already leads to a state in which a population with only cooperators becomes a Nash equilibrium, in addition to the already existing Nash equilibrium at a finite fraction of cooperators, which we observed in section \ref{sec:HDexpval}. As $\alpha$ is increased, the Nash equilibrium with $c=1$ becomes the only Nash equilibrium.
The phase diagram in figure \ref{fig:HDphasediagram} on the left shows the number of Nash equilibria for the HD game as a function of $\alpha$ and $\Delta t$ for scenario (a). The blue region, where only the Nash equilibrium at a finite fraction of cooperators increases in the late time limit, as expected from the analysis in section \ref{sec:HDexpval}.  

As is the case of the PD game, the appearance of the second Nash equilibrium for $c=1$ is a feature of the sharing of profits among players with like strategies combined with the risk aversion utility. If we instead consider scenario (c) then we do not see the appearance of this new Nash equilibrium but only the other Nash equilibrium at a non-zero fraction of c. The value of $c$ where this Nash equilibrium sits does change as a function of $\alpha$ and $\Delta t$, moving towards $c=1$ as $\alpha$ becomes larger and $\Delta t$ becomes smaller. In this situation, even though the second Nash equilibria does not appear, there is still a sense in which risk aversion can lead to cooperation as the Nash equilibrium moves towards $c=1$.

Interestingly, in scenario (b), the middle region of the phase diagram on the left of figure \ref{fig:HDphasediagram} where both Nash equilibria are present disappears. In this case, the location of the Nash equilibrium at a finite fraction of $c$ is moving towards $c=1$ as $\alpha$ is increased and there is no region where both Nash equilibria are present. The phase diagram for this scenario is shown in the right panel of figure \ref{fig:HDphasediagram}.  
In this case, the solution without risk aversion is the one where some part of the population cooperates to ensure that profit is gained, while the rest of the population free rides on their cooperation. Here, the risk experienced by the defectors at the classical Nash equilibrium has two sources: (1) as before they may not be selected to play the game, in which case they do not obtain profit; (2) the population at the classical Nash equilibrium may contain many defectors. This means that even if they are selected, they may earn the lowest possible profit by being paired with another defector. If the defectors share their profits, they can mitigate the risk coming from both of these sources. 

We conclude this section by noting that the risk aversion utility function of Definition \ref{def:RAutility} leads to cooperative behaviour in all the games we studied in this paper. In this section, we improve our analysis by performing stochastic simulations and determining the precise value of $p_X^{\rm min}(c)$ for a given $\alpha$ for the different games.

\section{Microscopic stochastic simulations}
\label{sec:simulation}
In this section, we study the reaction network \eqref{eq:genprionsersreactions} at the microscopic level using stochastic simulations. We will precisely recover several of the observables that we computed using the mesoscopic stochastic model of section \eqref{sec:stochasticmodel} once performing an ensemble average over trajectories. Besides showing the consistency of the picture depicted in figure \ref{fig:levels} we will also determine precisely the exact value of $p_X^{\rm min}(c)$ for the different games and corroborate the emergence of the new Nash equilibria that appeared in section \ref{sec:riskaversion}. We begin by detailing the simulation algorithm used before studying the reaction network in detail.

\subsection{Simulation algorithm}
\label{sec:simalg}
One of the advantages of framing population games as chemical reaction networks is that it is possible to export all the numerical machinery developed for chemical reaction simulations to population games. In the case of deterministic models, where one deals with simple ordinary differential equations (ODEs), one can simply use all the well-known ODE solvers \cite{leveque2007finite}. For stochastic models, the main algorithm to simulate stochastic trajectories of the underlying process is called the Gillespie algorithm or the stochastic simulation algorithm (SSA) \cite{anderson2015stochastic,gillespie1977exact}. 

Given an initial condition, the Gillespie algorithm applied to \cref{eq:genprionsersreactions} propagates the copy number of chemical species in time, where changes in the copy numbers are driven by reaction events. In the context of this work, instead of chemical species, we have players and gains and instead of reaction events, we have games to play. The essence of the algorithm consists of identifying two stochastic quantities: (1) the time for the next reaction (or game) to happen, $\tau_1$, and (2) which reaction (or game) will happen, denoted by the index $i$ in $R_i$. With these quantities, we can propagate the system forward in time by $\tau_1$, and we modify the copy numbers according to the ones specified by reaction (or game) $R_i$. For details on how to implement the Gillespie algorithm, we point the reader to \cite{anderson2015stochastic,gillespie1977exact}. 

The Gillespie algorithm is closely connected to the master equation (see \cref{eq:mastereqn}). The ensemble average of $\mathcal{N}$ realisations simulated with the Gillespie algorithm satisfy the probability distribution from \cref{eq:mastereqn} in the limit of $\mathcal{N}\rightarrow \infty$. This has been proved mathematically using the law of large numbers \cite{anderson2015stochastic}. We note, however, that given that the systems we study here are ergodic, the ensemble average over trajectories in the limit $\mathcal{N}\to\infty$ is equivalent to the average over a single trajectory in the limit $T\to\infty$ where $T$ is simulation time. We will first show that some of the results obtained using the stochastic model \eqref{sec:stochasticmodel} can be obtained using stochastic simulations.

\subsection{Recovering the stochastic model via ensemble averaging}
\begin{figure}[h!]
    \centering
    \includegraphics[scale=0.6]{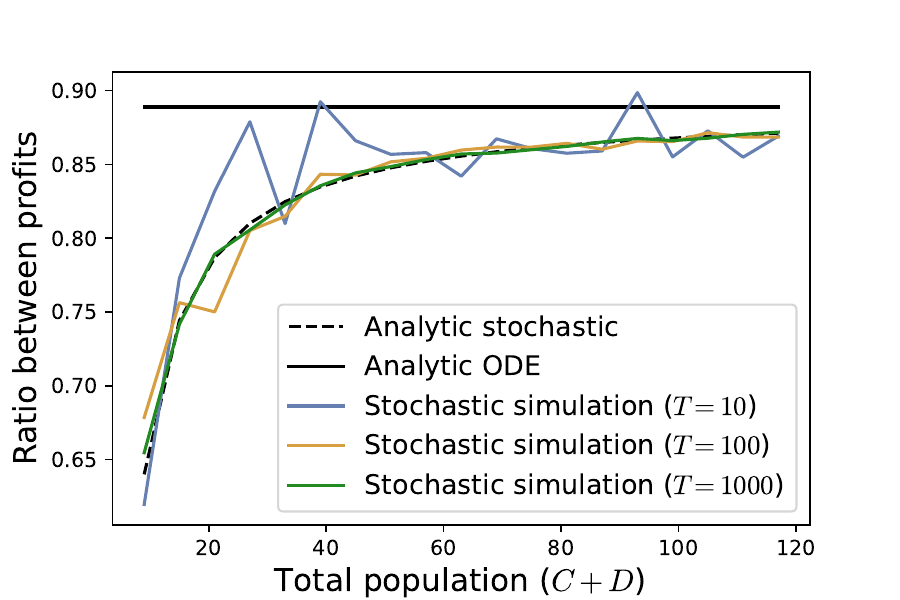}
    \caption{Profit ratio $p_r$ convergence for PD game with $\tau=1$, $r=0.8$, $p=0.2$, $s=0$ as a function of total population size $N=C+D$. The analytical profit ratios for the deterministic and the stochastic case are the ones given in \cref{profitratio,stochprofratio}, respectively. The profit ratio for the stochastic simulations are calculated at three final times, $T=10,100,1000$.}
    \label{fig:profitRatioConvergence}
\end{figure}

\begin{figure}[h!]
    \centering
    \includegraphics[scale=0.29]{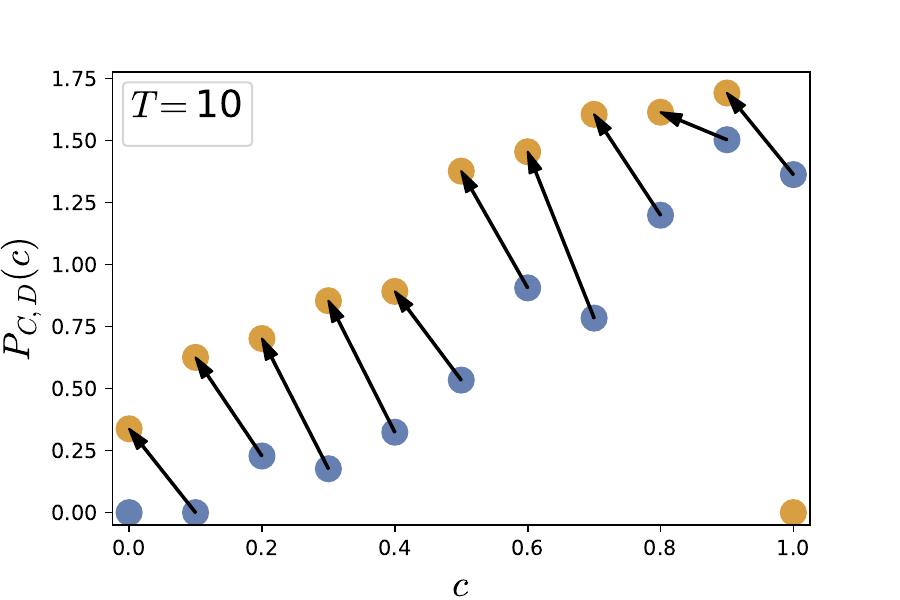} 
    \includegraphics[scale=0.29]{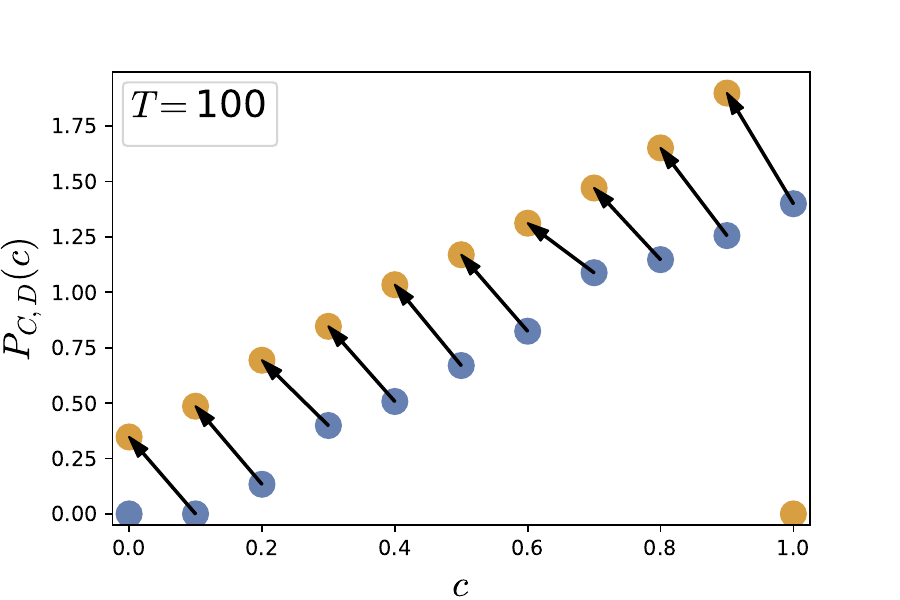}
    \includegraphics[scale=0.29]{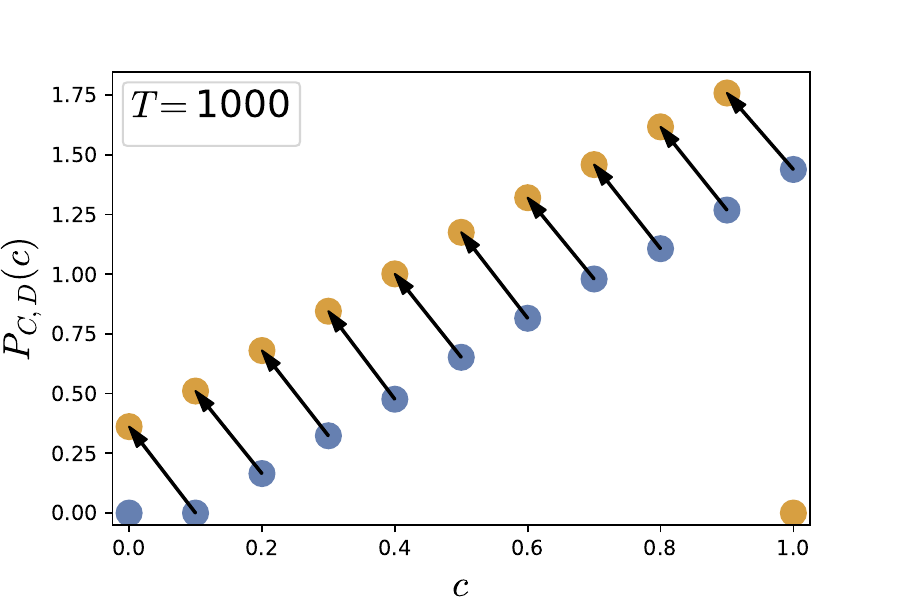}
    \caption{Profit rate of the PD game for defectors (yellow) and cooperators (blue) in a population of $N=10$ as a function of the population composition $c$. These figures were calculated using the stochastic model at three final times, $T=10,100,1000$ with $\tau=1$, $r=0.8$, $p=0.2$, $s=0$.}
    \label{fig:profitRatioPDstoch}
\end{figure}

In order to show that we can indeed recover the stochastic model of section \ref{sec:stochasticmodel} using this microscopic level of description, we use the Gillespie algorithm to simulate realisations of the PD game. In figure \ref{fig:profitRatioConvergence}, we show the analytic results of the profit ratios obtained from the macroscopic level of description in \cref{profitratio} and the profit ratios obtained using the stochastic model in \eqref{stochprofratio}. 
In addition, we also plot in figure \ref{fig:profitRatioConvergence} the profit ratios obtained from three stochastic simulations at three different final times. As the population size $N=C+D$ grows, one can clearly observe that the expected value of the stochastic profit rate tends towards the macroscopic value. Moreover, the stochastic simulations approach the expected value obtained using the stochastic model as the profit ratio is calculated over a longer period of time. This is due to the fact that as the gain increases over time, the fluctuations remain of equal magnitude and thus the relative fluctuations decrease to zero. This is a consequence of the law of large numbers. We can also determine the profit rates as a function of the population composition which we depict in \cref{fig:profitRatioPDstoch}. 
As shown in \cref{fig:profitRatioConvergence}, the final time at which the profit rates are calculated influences the amount of fluctuations in the system. We see the same pattern in \cref{fig:profitRatioPDstoch}, in particular as the time $T$ increases, the fluctuations become more and more negligible, and we find an exact match with the results in figure \ref{fig:NEPrisonersDilemma}. This comparison is important as it reveals the consistency of the picture that we described in figure \ref{fig:levels}. 

It is straightforward to extend these simulations to the SH and HD games by simply changing the model parameters. Instead, we will now use this method to determine numerically the Nash equilibria for the different games using the risk aversion utility.

\subsection{Numerical determination of Nash equilibria}
\begin{figure}[h!]
    \centering
    \includegraphics[scale=0.29]{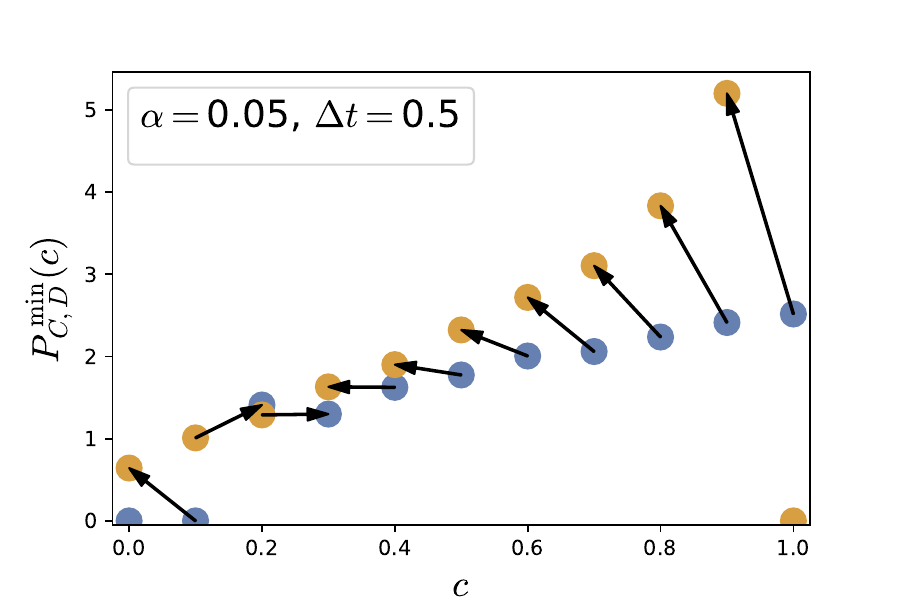} 
    \includegraphics[scale=0.29]{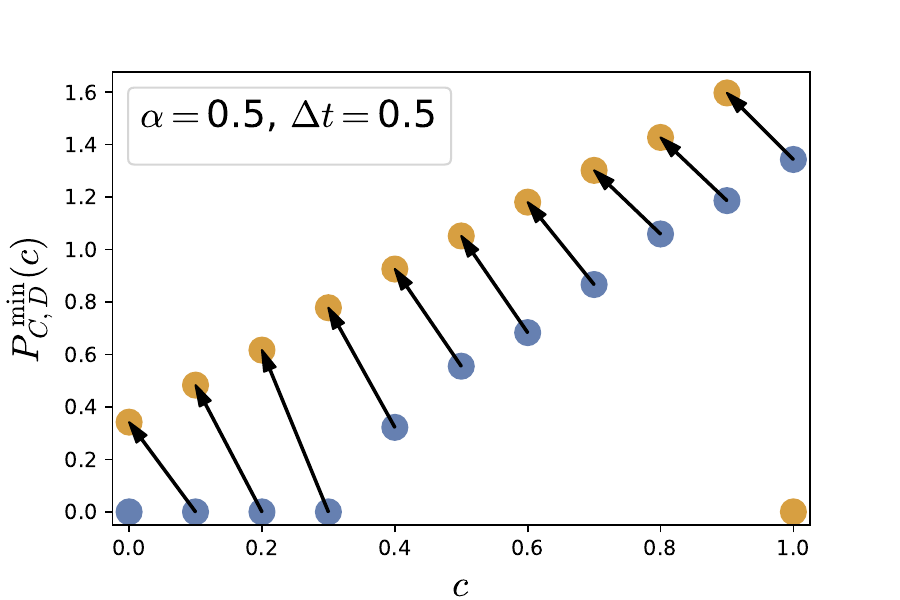}
    \includegraphics[scale=0.29]{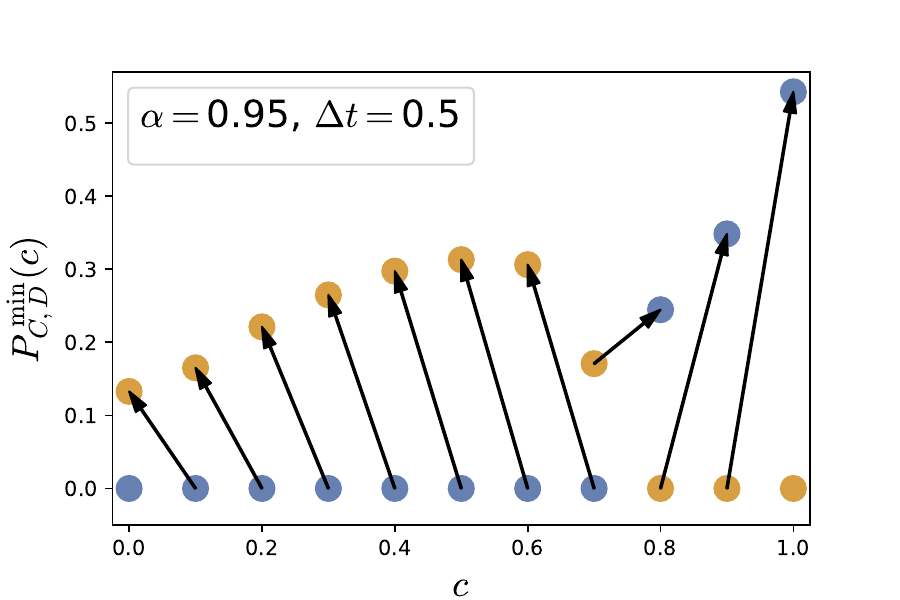} \\
    \includegraphics[scale=0.29]{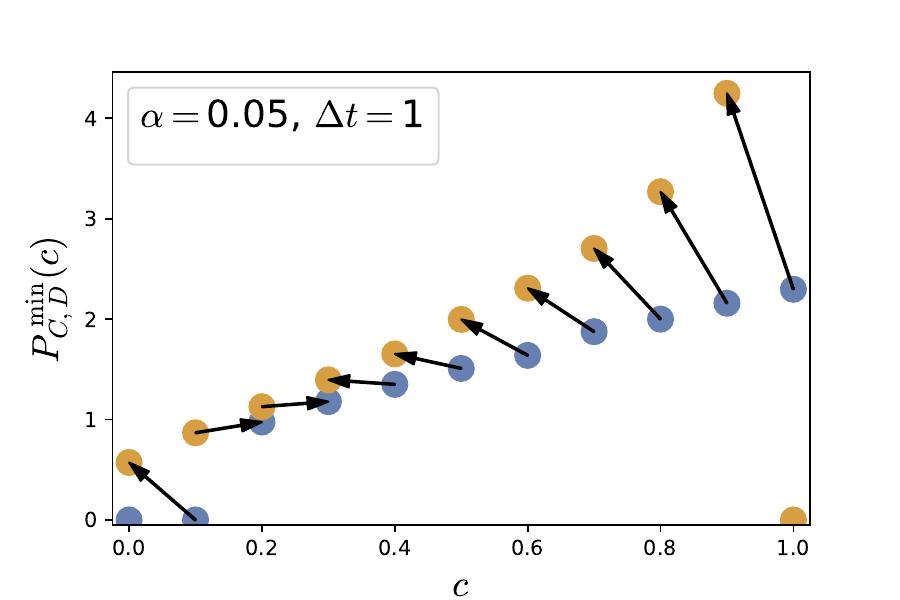}
    \includegraphics[scale=0.29]{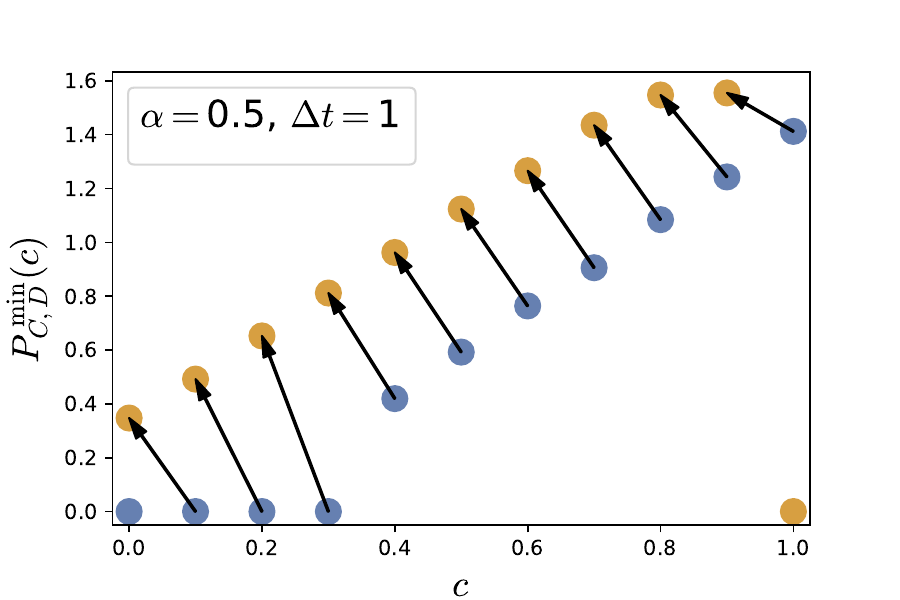}
    \includegraphics[scale=0.29]{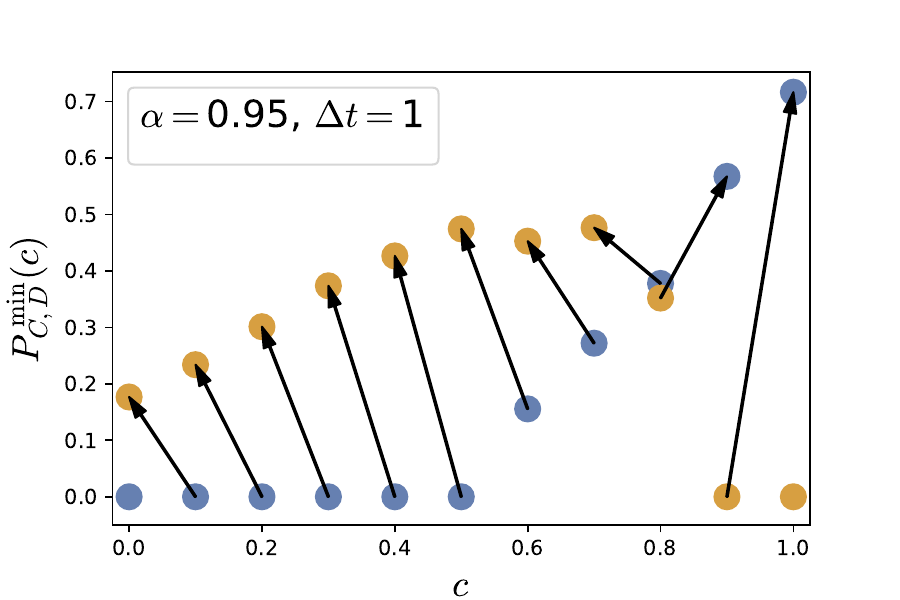} \\
    \includegraphics[scale=0.29]{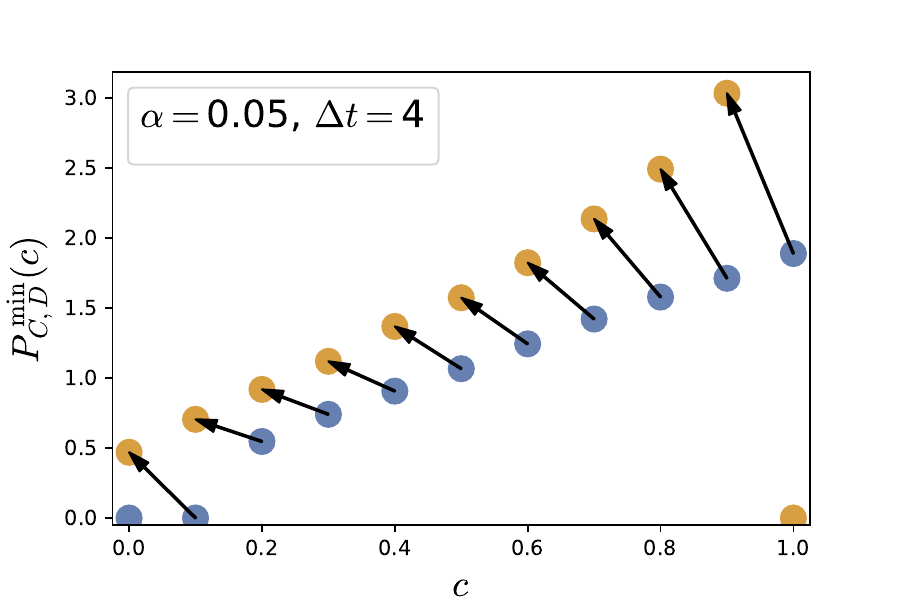}
    \includegraphics[scale=0.29]{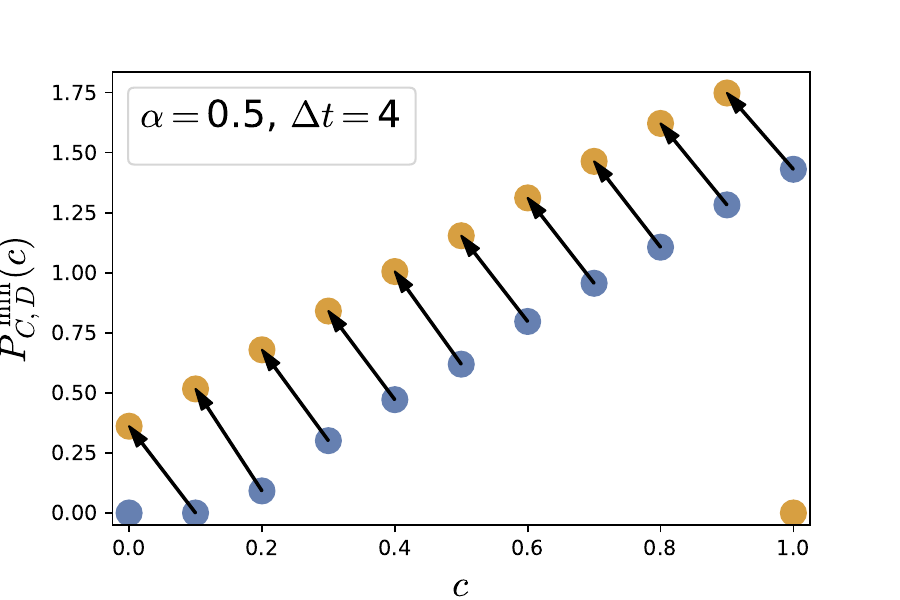}
    \includegraphics[scale=0.29]{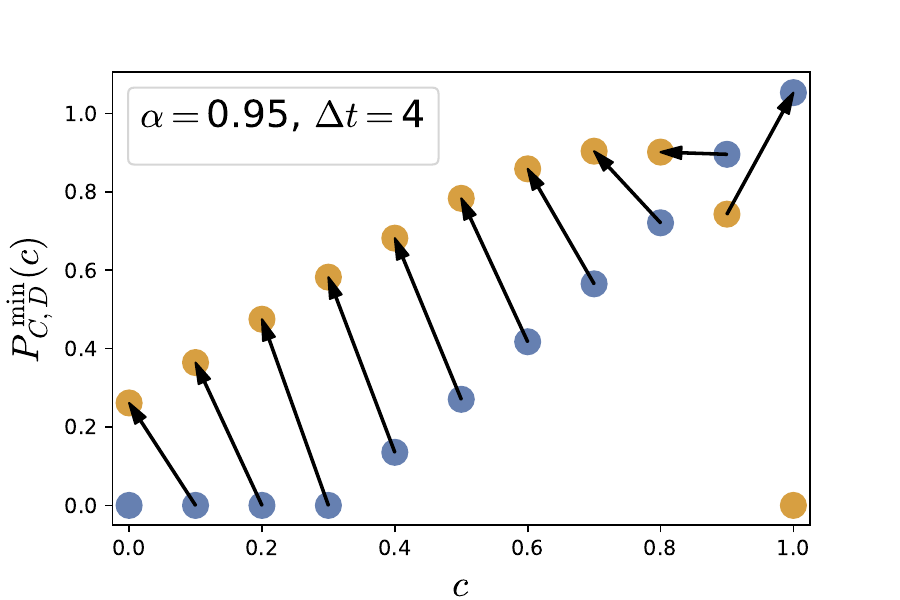} \\
    \caption{Minimum payout of the PD game with risk aversion for defectors (yellow) and cooperators (blue) in a population of $N=10$. These figures were obtained using stochastic simulations with final time $T=3000$ for different values of risk aversion $\alpha$, where the risk aversion was evaluated over a time interval of $\Delta t$ and averaged over $\mathcal{N}=1500$ realizations. The first row corresponds to $\Delta t =0.5$, the second row to $\Delta t =1$ and the third row to $\Delta t=4$. The columns correspond to $\alpha=0.05,0.5,0.95$, respectively. }
    \label{fig:profitRatioPD_risk_stoch}
\end{figure}

\begin{figure}[h!]
    \centering
    \includegraphics[scale=0.29]{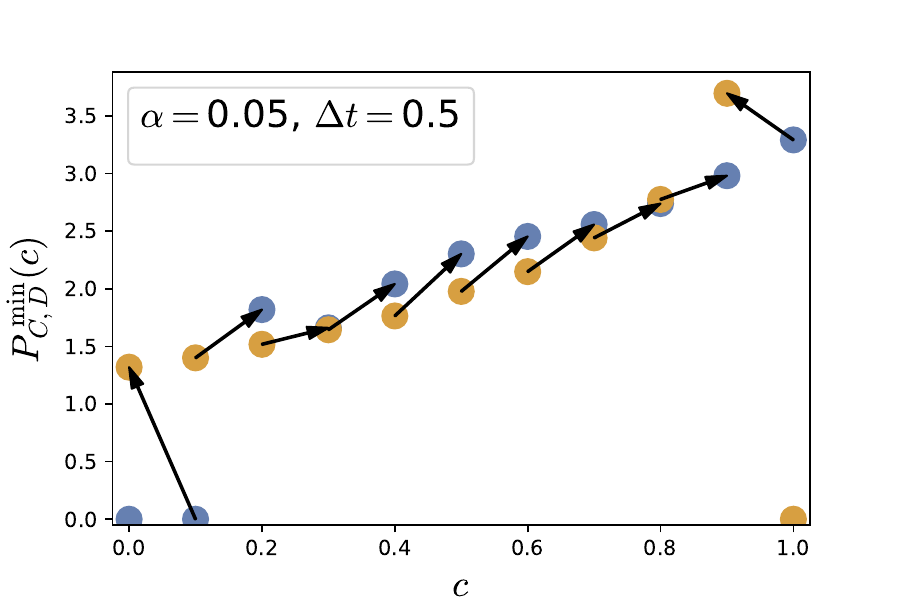} 
    \includegraphics[scale=0.29]{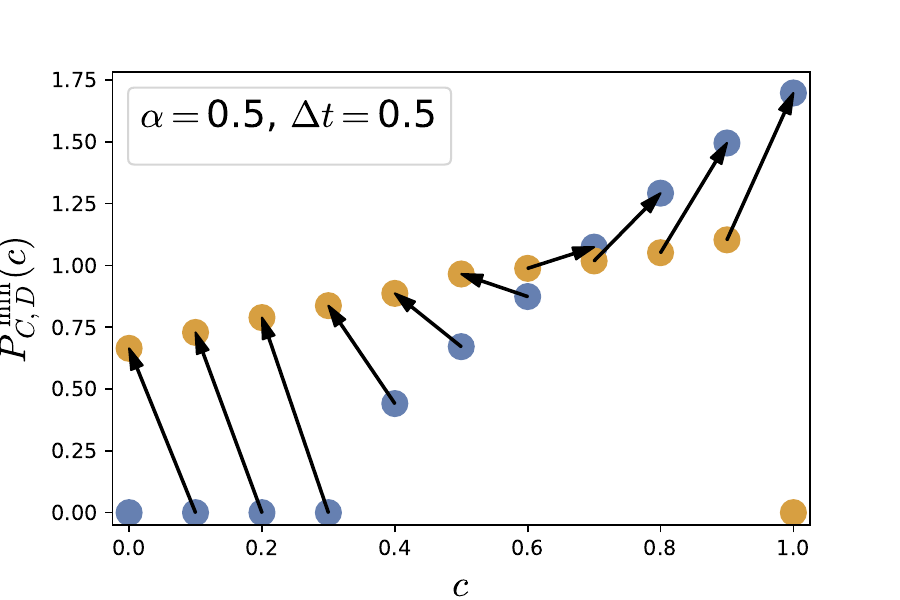}
    \includegraphics[scale=0.29]{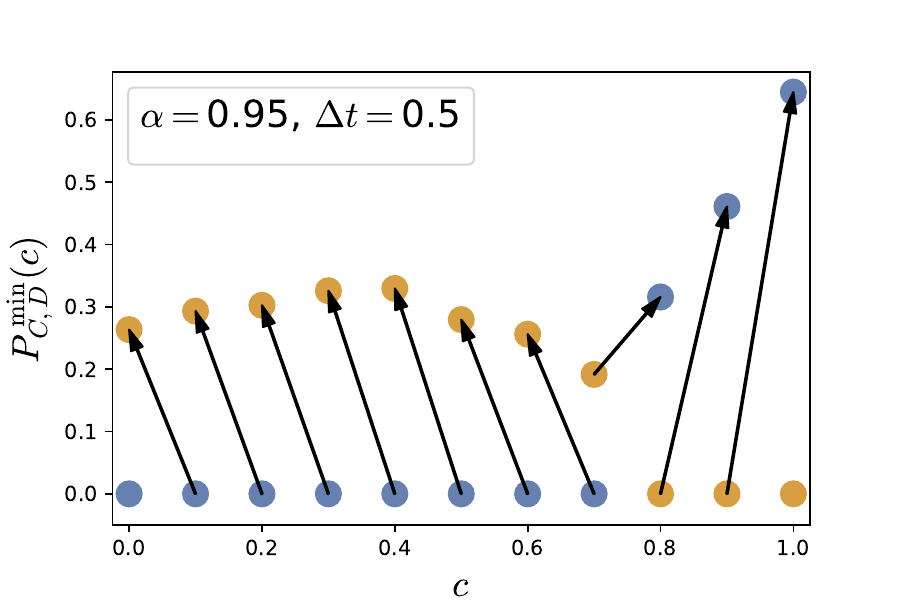} \\
    \includegraphics[scale=0.29]{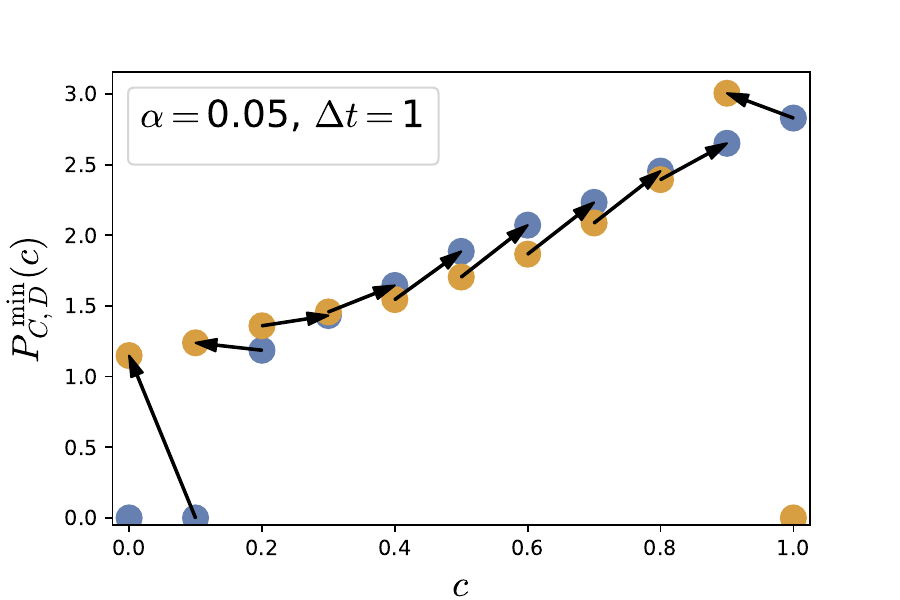}
    \includegraphics[scale=0.29]{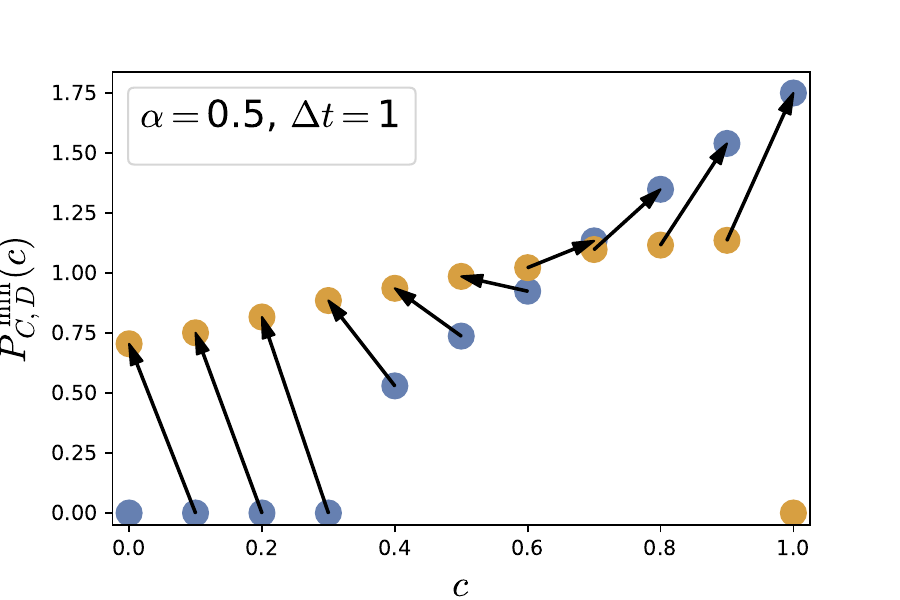}
    \includegraphics[scale=0.29]{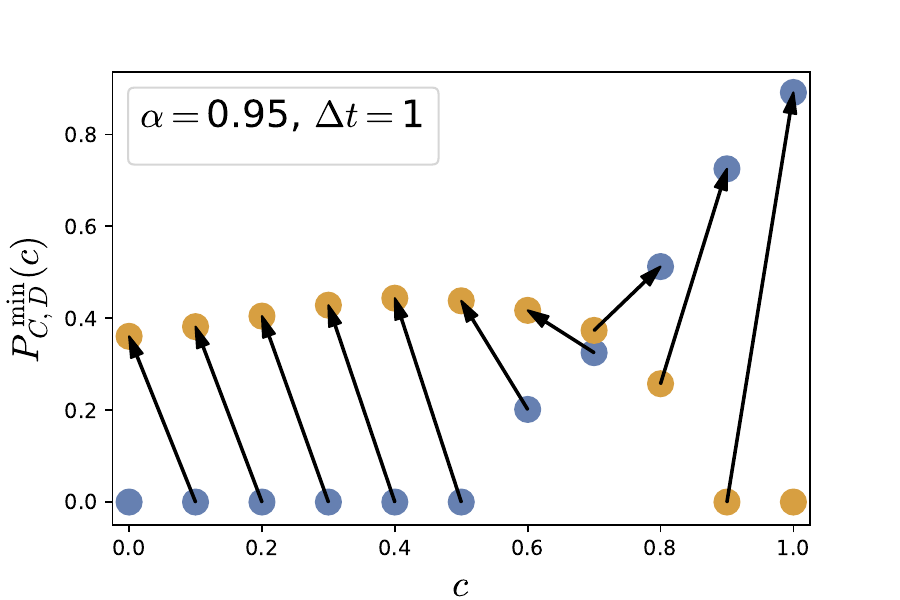} \\
    \includegraphics[scale=0.29]{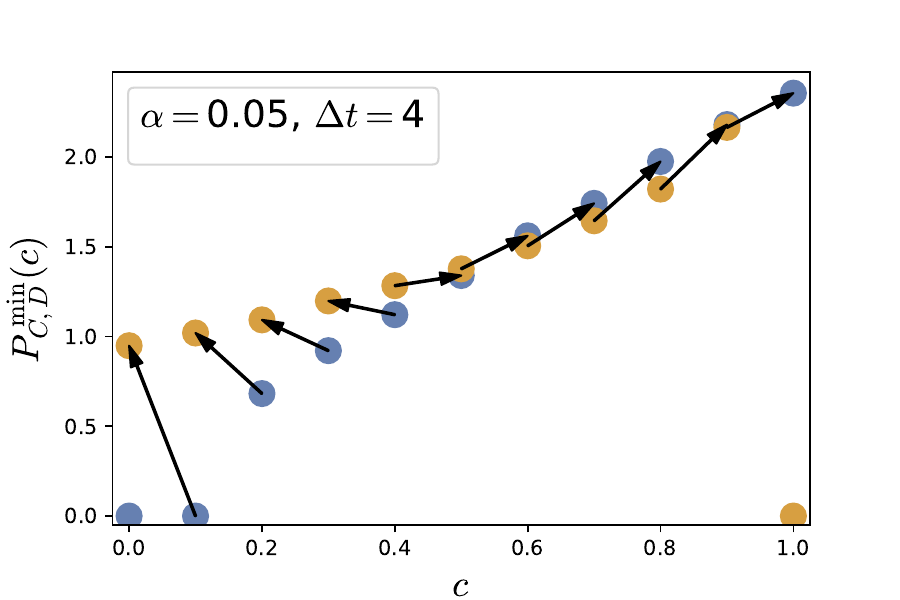}
    \includegraphics[scale=0.29]{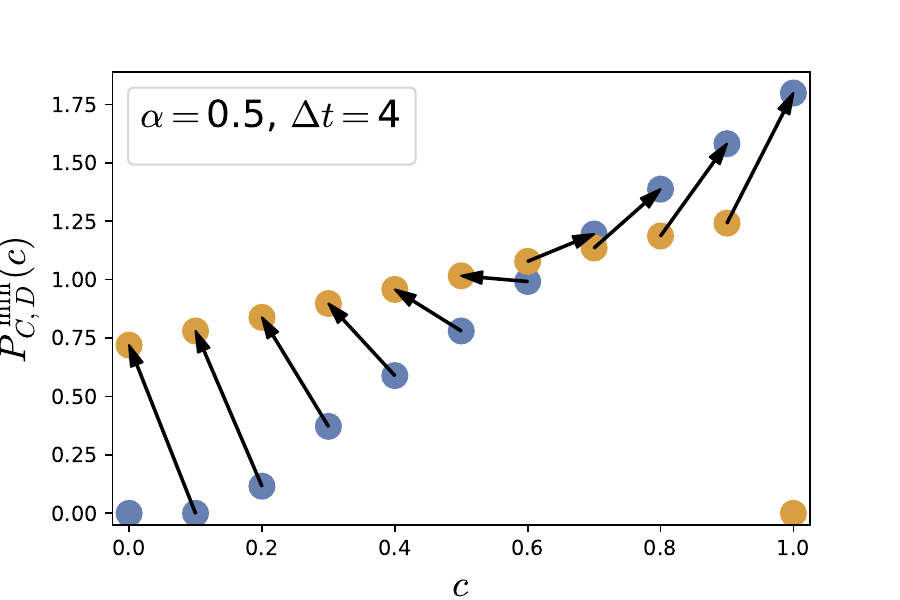}
    \includegraphics[scale=0.29]{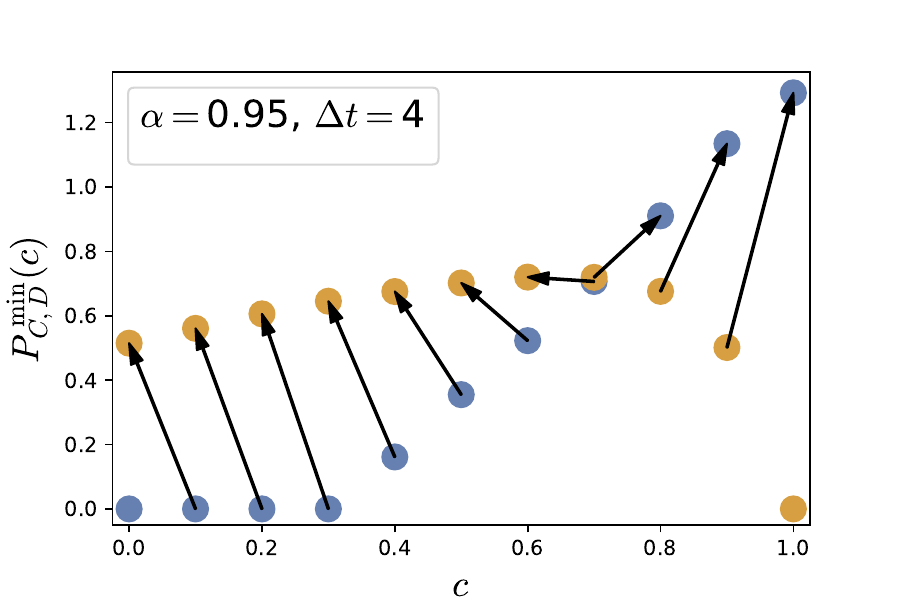} \\
    \caption{Minimum payout of the SH game with risk aversion for defectors (yellow) and cooperators (blue) in a population of $N=10$. These figures were obtained using stochastic simulations with final time $T=3000$ for different values of risk aversion $\alpha$, where the risk aversion was evaluated over a time interval of $\Delta t$ and averaged over $\mathcal{N}=1500$ realizations. The first row corresponds to $\Delta t =0.5$, the second row to $\Delta t =1$ and the third row to $\Delta t=4$. The columns correspond to $\alpha=0.05,0.5,0.95$, respectively.}
    \label{fig:profitRatioSH_risk_stoch}
\end{figure}

\begin{figure}[h!]
    \centering
    \includegraphics[scale=0.29]{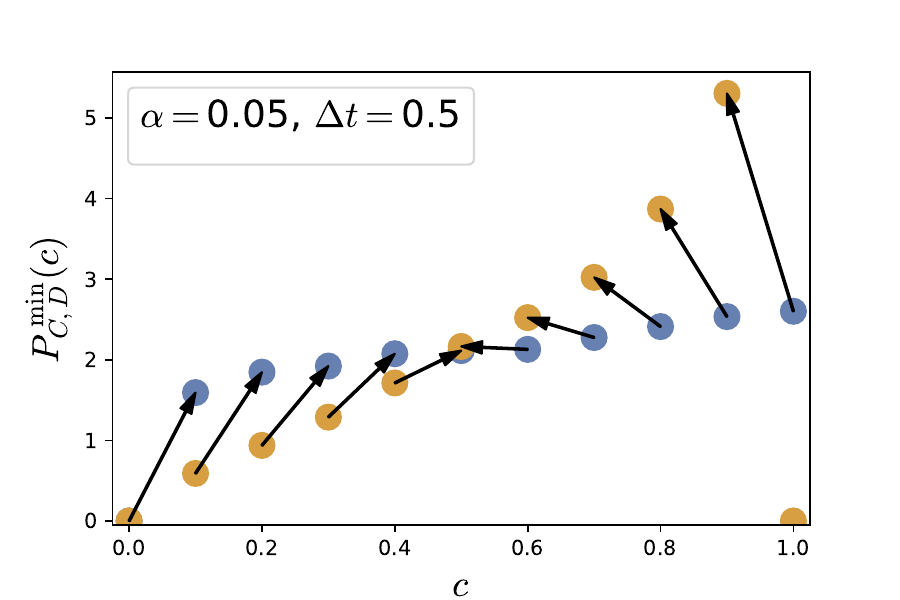} 
    \includegraphics[scale=0.29]{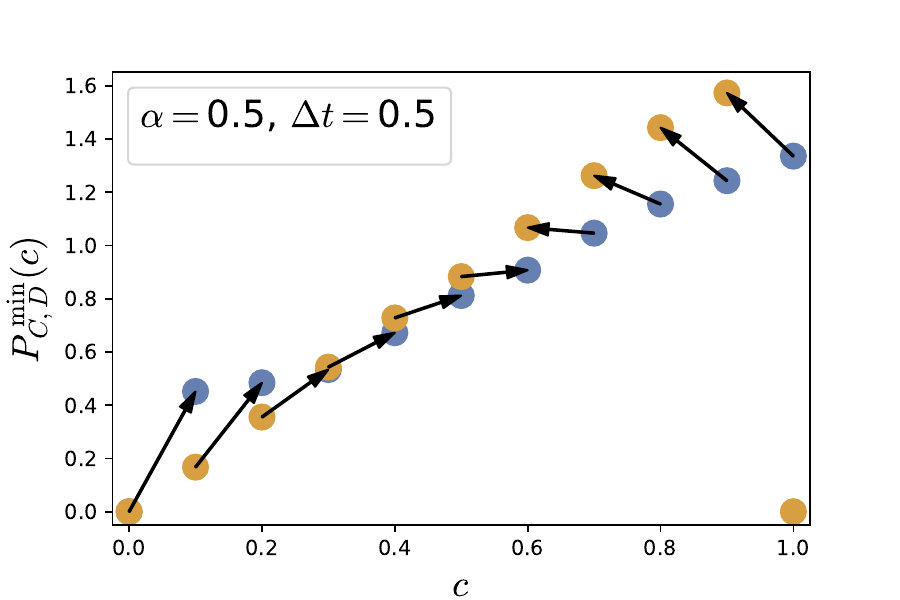}
    \includegraphics[scale=0.29]{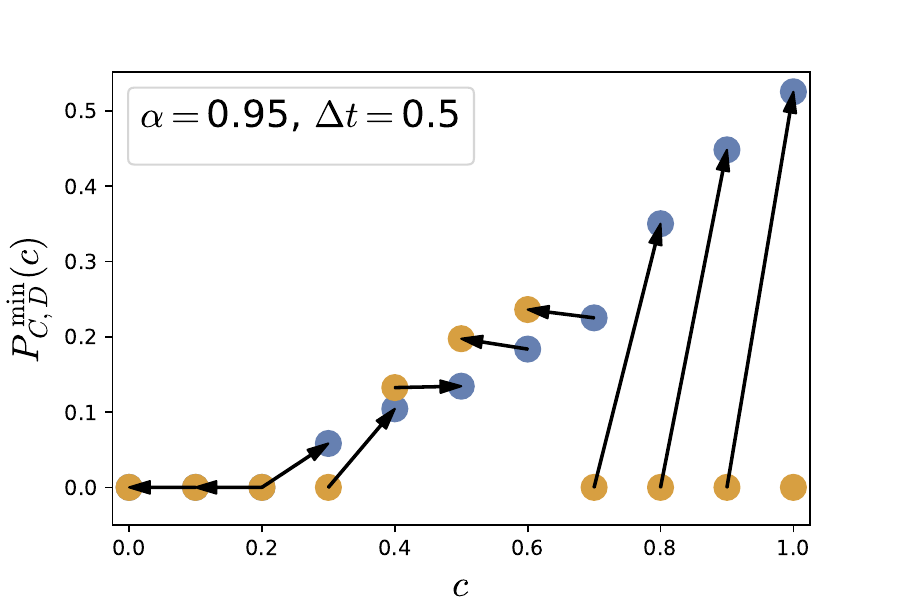} \\
    \includegraphics[scale=0.29]{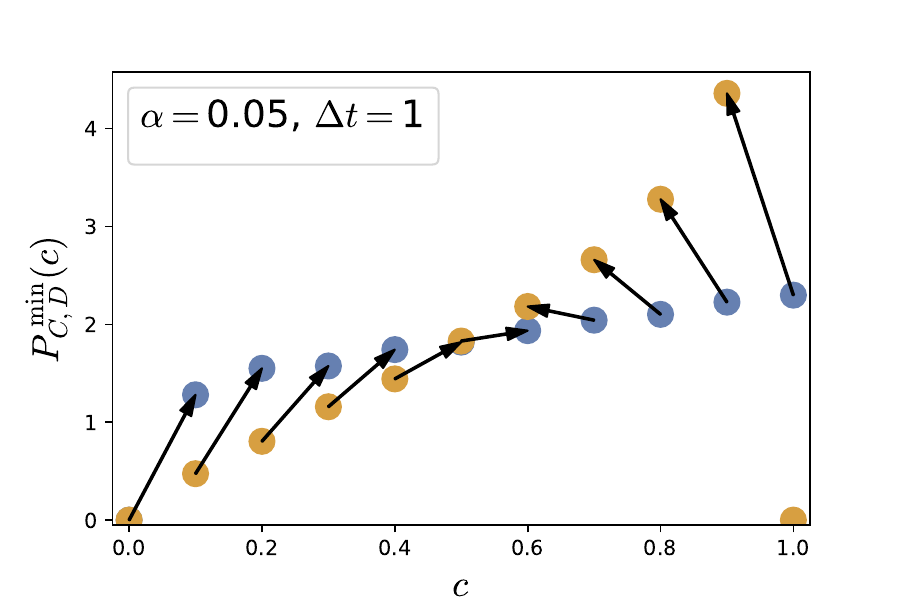}
    \includegraphics[scale=0.29]{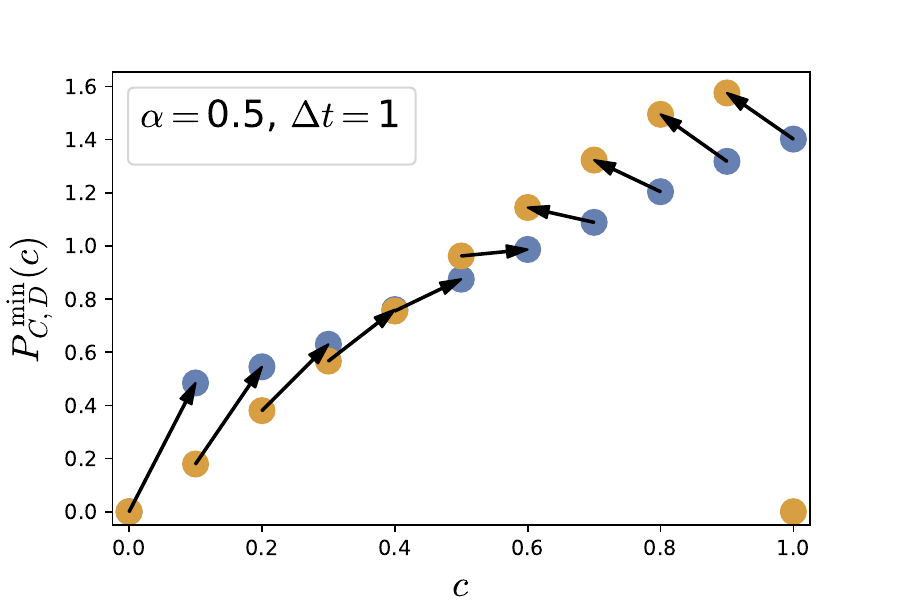}
    \includegraphics[scale=0.29]{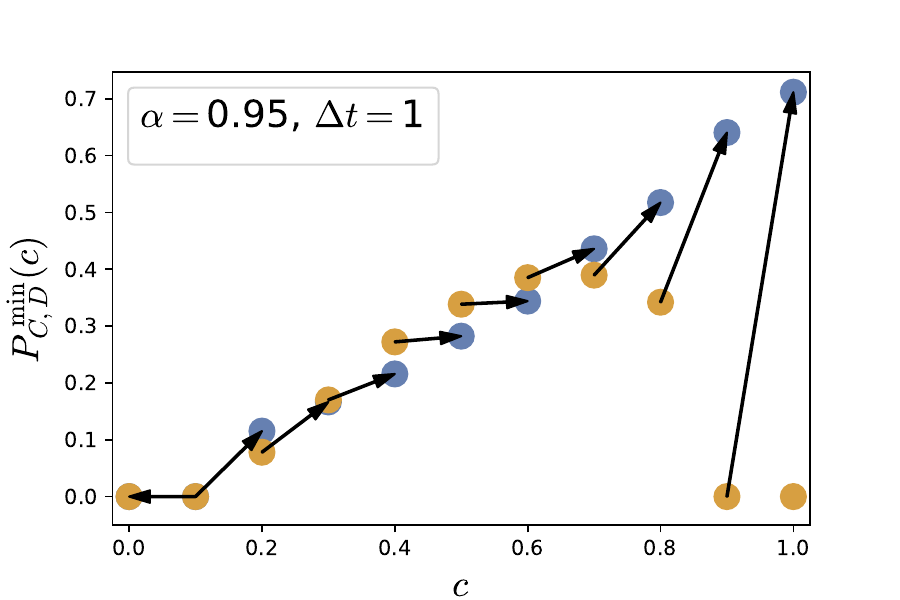} \\
    \includegraphics[scale=0.29]{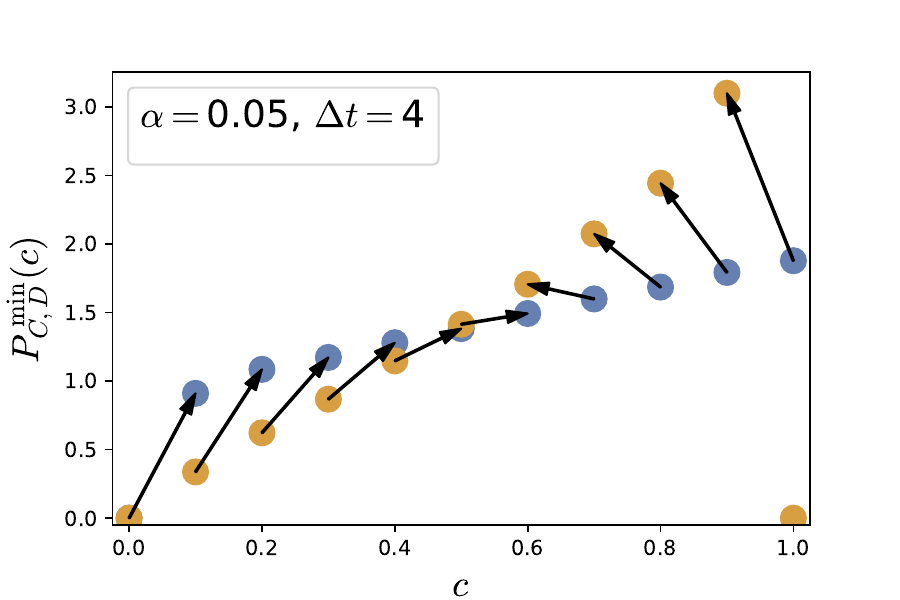}
    \includegraphics[scale=0.29]{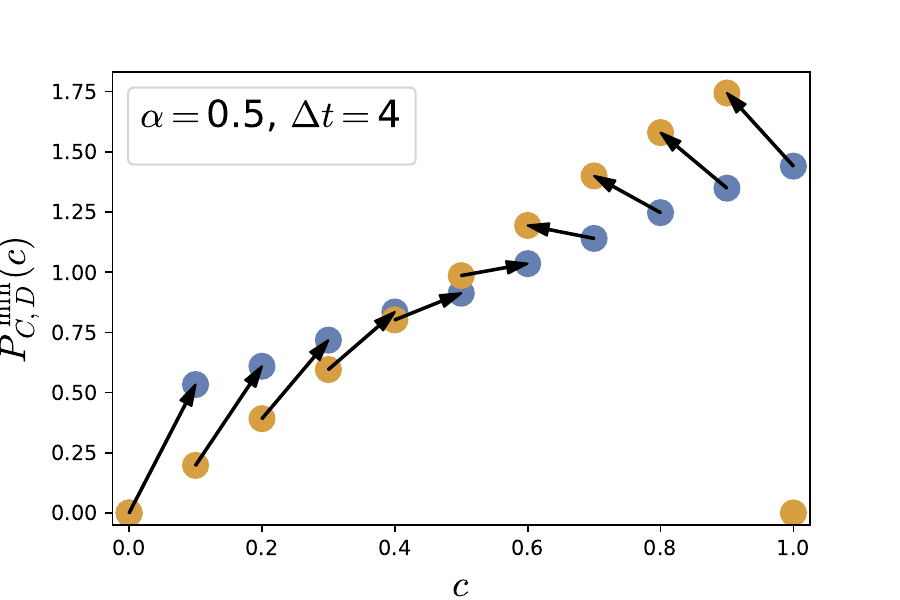}
    \includegraphics[scale=0.29]{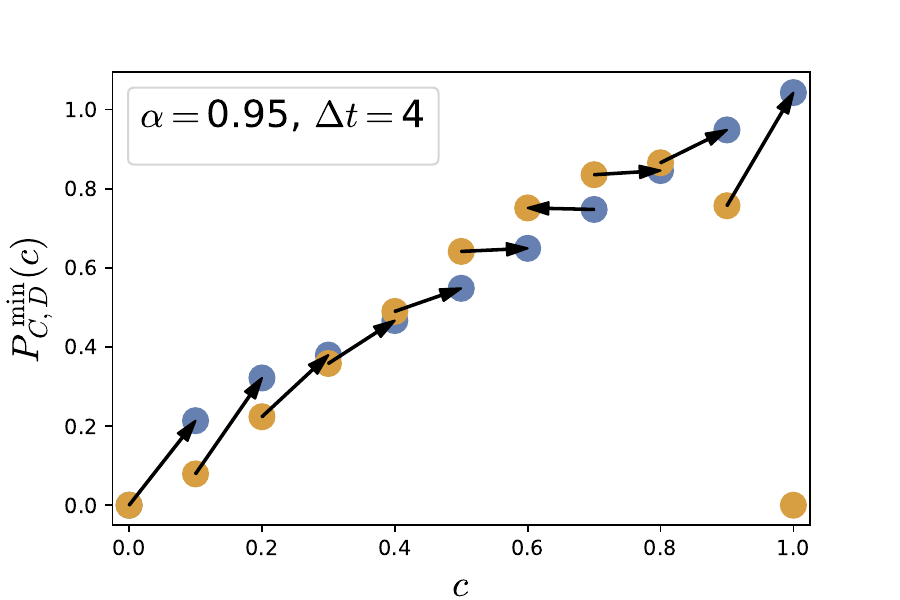} \\
    \caption{Minimum payout of the HD game with risk aversion for defectors (yellow) and cooperators (blue) in a population of $N=10$. These figures were obtained using stochastic simulations with final time $T=3000$ for different values of risk aversion $\alpha$, where the risk aversion was evaluated over a time interval of $\Delta t$ and averaged over $\mathcal{N}=1500$ realizations. The first row corresponds to $\Delta t =0.5$, the second row to $\Delta t =1$ and the third row to $\Delta t=4$. The columns corresponds to $\alpha=0.05,0.5,0.95$, respectively.}
    \label{fig:profitRatioHD_risk_stoch}
\end{figure}

In this section, we delve into the risk-averse case using numerical simulations. Our main goal is to determine the Nash equilibria using the risk aversion utility function for the different games of section \ref{sec:riskaversion}. In particular, we want to numerically determine the minimum payout $p_X^{\rm min}(c)$ for given confidence levels $\alpha$ and verify the emergence of cooperation in these models. The risk aversion confidence level $\alpha$ is evaluated by averaging over a time interval 
$\Delta t$. We incorporate this into the simulation algorithms presented in \cref{sec:simalg}. We then perform simulations for the risk-averse case for the PD, SH and HD games, and we show the risk-averse utilities obtained for different values of $\alpha$ and $\Delta t$ in \cref{fig:profitRatioPD_risk_stoch,fig:profitRatioSH_risk_stoch,fig:profitRatioHD_risk_stoch}, respectively for each game. 

In the three games investigated numerically in \cref{fig:profitRatioPD_risk_stoch,fig:profitRatioSH_risk_stoch,fig:profitRatioHD_risk_stoch}, we observe that a higher level of risk aversion (larger $\alpha$) consistently leads to cooperation. We also observe non-trivial emergence of Nash equilibria depending on the degree of risk aversion $\alpha$ as well as on the time interval $\Delta t$ used to average it. For instance, in \cref{fig:profitRatioHD_risk_stoch} for $\alpha=0.95$ and $\Delta t = 4.0$, we note that there is one Nash equilibrium at $c=1$ and another one around $c=0.6$. As we decrease the time interval to $\Delta t=1$, one Nash equilibria disappears while another Nash equilibrium emerges at $c=0$, and similarly for  $\Delta t=0.5$, another non-trivial  Nash equilibrium emerges at around $c=0.4$. All the results of exhibited in \cref{fig:profitRatioPD_risk_stoch,fig:profitRatioSH_risk_stoch,fig:profitRatioHD_risk_stoch} match the corresponding results in section \ref{sec:riskaversion} which were based on the lower bound for the risk utility, confirming the existence of the novel Nash equilibria. We also note that we have only presented numerical simulations for scenario (a) but we are confident that the same match would be obtained if we would perform simulations for scenarios (b) and (c). 

The numerical framework used here provides a robust approach to simulate cooperation games, including risk-averse cases, and it can be extended to handle more complex settings, e.g. alternative utilities measures, strategic learning, network-dependent dynamics or spatial-dependent dynamics. In addition, we showed that it can be used to obtain results that are difficult or not possible to attain analytically, such as the minimum payout $p_X^{\rm min}(c)$. The simulations also enable us to determine the Nash equilibria, the long time dynamics, as well as the ensemble behaviour of large populations and their characteristic fluctuations.

\section{Conclusion}
\label{sec:conclusion}
What is the mechanism driving the evolution of cooperation? And if we may speculate further, what mechanism underlies the evolution of living systems? The majority of works have addressed this question from the point of view of maximisation of individual or group fitness, with the purpose of obtaining as much benefit or profit as possible in the long run. Yet, there is no conclusive mechanism that is able to explain the emergence of cooperation across contexts, across length scales, across species and across group sizes. In this paper, we have challenged this common assumption and proposed a novel mechanism for the evolution of cooperation based on the notion of a risk-averse utility function. Instead of seeking to maximise average benefits in the infinite future, individuals seek to guarantee a certain minimum of resources with a given probability and in a given time span. This utility function has several interesting properties and gives rise to many new results that we now summarise:
\begin{itemize}
\item The risk-averse utility function is dependent on the particular time span $\Delta t$ that groups or individuals wish to maximise their minimum profit rates with a given level of confidence. This notion seems to be in line with the fact that living systems have limited lifetimes and want to guarantee their survival in the next stage of the game or the next stage of development. 
\item The risk-averse utility function is continuously connected to the utility based on average profit rates in the limit $\Delta t\to\infty$. This means that if there are systems or living organisms that at some point in time or in stage of development are in fact driven by average profit maximisation, the utility function we proposed is also applicable to those cases. 
\item Maximising the risk-averse utility leads to new Nash equilibria and to cooperative behaviour in all the cooperation games that we considered, in particular prisoner's dilemma, stag hunt, and hawk-dove games. This means that the mechanism we proposed is able to explain the emergence of cooperation for a wide range of potential interaction dynamics. Very importantly, we did not bias the system by introducing any type of selection or population structure (e.g. by tuning reaction rates to be comparatively large in some reactions) or any type of reciprocity mechanism (e.g. retaliation or sanctions). In this sense, the emergence of new Nash equilibria that promote cooperation is a robust result for which we expect that only by strongly biasing the system can they be removed. 
\item An essential ingredient for emergence of cooperation in our setting is the combination of risk aversion utility with the sharing of profits among individuals of the same type. Interestingly, scenario (b), in which only the cooperators share their profits, is sufficient for observing the second Nash equilibrium. 
\item We have particularly shown that cooperative behaviour can be achieved in any population with $N\ge3$. This means that the mechanism we proposed is scalable to any population size. 
\end{itemize}
While we are confident of the value of this new principle driving the evolution of cooperation based on risk aversion, it would be very interesting to understand how robust the mechanism is. In particular, it would be important to test whether these new Nash equilibria are stable against different sources of noise in the system. Specifically, we may introduce random mutations that can swap a cooperator for a defector, and vice versa; random fluctuations in the payout parameters modelling the possibility that not all interactions between every member of the groups are necessarily the same\footnote{Introducing fluctuations of the off-diagonal components of the payout matrix (see the table in section \ref{sec:cooperationgames}) can model heterogeneity of \emph{risk perception} in certain social contexts (see e.g. \cite{PhysRevE.99.052311, doi:10.1098/rspa.2020.0116, doi:10.1098/rspa.2020.0116}), that is the perception of players regarding their potential temptation or sucker's payout. This is unrelated to the \emph{risk aversion utility function} that we introduced here. It would nevertheless be interesting to understand the effect of such fluctuations in the context of risk aversion utility.}; spatial structure, network structure and spatial dynamics to test if and how cooperative behaviour can spread in various social and economic contexts. The methods we employed here can be extended to address these cases.

It is important to stress that in this paper we have introduced a novel and robust framework for studying cooperative games with arbitrary large groups of populations. This framework moves beyond the modelling of cooperation games using macroscopic rate equations for chemical reaction networks proposed in \cite{veloz2014reaction}, which is only valid for a very large number of players, by considering more fundamental levels of description. In particular, as in the context of bio-molecular reactions in which reaction networks such as \eqref{eq:genprionsersreactions} are commonplace, we describe the microscopic dynamics of the system using stochastic modelling valid for any number of players. The stochastic model of cooperation games is best formulated using the language of quantum mechanics, in which the appearance of new \emph{species} (i.e. players or gains) can be understood as acting on probability states with \emph{raising operators}. Exploring such techniques allows us to extract a lot of analytic information about the system, including the moment generation function which contains all statistical information about the system, and to define novel measures of fitness that had not been considered earlier in the literature. Furthermore, we make use of suitable numerical schemes for solving molecular kinetics in order to perform stochastic simulations of cooperation games. We show that such simulations are consistent with the analytic stochastic model we employ, and in addition allow us to extract specific information about the system that is hard to obtain analytically. We believe that the combination of these methodologies provides a promising framework for studying population games which is generically applicable to a wide range of contexts. In this paper we only scratched the surface both at the level of specific systems but also at the level of methodologies. Indeed, an interesting methodology that would enrich this framework even further is large deviation theory \cite{touchette2012basic}, which would allow us to, for instance, perturbatively obtain the probability distribution $\mathbb{P}[P_X(c)]$ as well as obtain approximate analytic values for $p_X^{\rm min}$. We intend to incorporate such methodology in a forthcoming publication.

As we mentioned in section \ref{sec:introduction} exploring different forms of reciprocity by endowing players with memory and the possibility of changing strategies is relevant for small group sizes and in particular in the context of human interactions.  In such contexts, it would be interesting to revisit previous models that study one-period strategies and study whether the risk-averse utility is able to lead to new results for small groups. An example of such a strategy is the \emph{win-stay, lose-shift strategy}, recently shown to be an equilibrium strategy for all three of the games considered here \cite{Meylahn2021} and which is a driver of cooperation in the prisoner's dilemma as shown in \cite{BarfussMeylahn2023, meylahn2023weak}. It would be timely to understand whether the same equilibrium strategy is attained under risk aversion. In addition, and while not directly relevant for large groups, we can consider a model with both local and global interactions. At the local level, a few members could be tied together in a network where reciprocity or selection plays a role, while at the global scale only risk aversion is relevant. Models that couple local and global dynamics have been considered in the literature in the setting of public goods games \cite{novakgoods} and it would be interesting to understand if risk aversion is relevant in such scenarios. In fact, it would be natural to revisit the literature on public good games in general, as it is known that cooperative behaviour is extremely hard to achieve \cite{BERGSTROM198625, lineargoods}, and check whether risk 
aversion could have a positive effect. We leave some of these questions for future work.

\subsection*{Acknowledgments} 
We would like to thank Akash Jain for initial collaborations in this project. We also would like to thank Andr\'{e} de Roos, Han van der Maas, Marco Javarone, Matthijs van Veelen, Tomas Veloz and Simon Toussaint for useful discussions and feedback. JA is partly supported by the Dutch Institute for Emergent Phenomena (DIEP) via the programme Foundations and Applications of Emergence (FAEME). SR is fully supported by DIEP. WM is partly supported by DIEP. JM and MJR have been funded by DIEP during part of this work. MJR is also partly supported by DFG Grant No. RA 3601/1-1.


\bibliographystyle{elsarticle-harv} 
\bibliography{references}





% \end{thebibliography}

\appendix

\section{Derivation of the moment generating function for the profit rate}
\label{app:LDT}
In this appendix, we detail the derivation of the moment generating function for the profit rates $\hat{P}_X$ with $X \in \{C, D\}$. Recall that the profit rate operators are defined in terms of the number operators for the gains obtained by following strategy $X$ as
\begin{equation}\label{profitoperatorsAPP}
    \hat{P}_C = \frac{1}{t C_0} N_{G_c} \,, \qquad 
    \hat{P}_D = \frac{1}{t D_0} N_{G_d} \,.
\end{equation}
In general, the moment generating function for any number operator $N_X = a_X^\dagger a_X$ is given by
\begin{equation}
    M_X(s_x) = \langle e^{s_x N_X} \rangle = \left. e^{s_x N_X} e^{ H t} P(0,\vec{z}) \right|_{\vec{z} = \vec{1}} \,.
\end{equation}
As we will show here, this can be expressed in terms of the exponential of a tilted operator $\tilde{H}(s_X)$, obtained from the generator $H$ by replacing creation operators for the species $X$ with $a^{\dagger}_X \to e^{s_X}a^{\dagger}_X$ and annihilation operators with $a_X \to e^{-s_X} a_X$. In general, we have that
\begin{equation}
    M_X(s_X) = \left.  e^{\tilde{H}(s_X) t} e^{s_X N_X} P(0,\vec{z}) \right|_{\vec{z} = \vec{1}},
\end{equation}
with
\begin{equation}
    \tilde{H}(s_X) = H(a^\dagger_X \to e^{s_X} a^\dagger_X, a_X \to e^{-s_X} a_X)~.
\end{equation}
To prove this equation, we use a corollary of the Baker-Campbell-Hausdorff formula, which states that for non-commuting operators $A$ and $B$, namely
\begin{equation}
e^{A} e^B = \exp \left( \sum_{n=0} \frac{1}{n!} [(A)^n , B] \right) e^A~~,
\end{equation}
where $[(A)^n, B]$ is shorthand notation for the nested commutator
\begin{equation}
[(A)^n, B] = \underbrace{[A, [\ldots , [A}_{n} ,B]]]~~.
\end{equation}
Applying this formula with $A = s_x N_X $ and $B = H t$, we obtain
\begin{equation}\label{ZNinter}
M_{X}(s_x, t) = \left. \exp \left(\sum_{n=0}^{\infty} \frac{s_x^n}{n!} [(N_X)^n, H] t \right) e^{s_x N_X} P(0,\vec{z}) \right|_{\vec{z} = \vec{1}} \,.
\end{equation}
To compute the nested commutators, one can use the commutation relations \eqref{comrel} to put the resulting terms back in normal ordering, where creation operators $a_X^\dagger$ are all to the left of the annihilation operators $a_X$. We start with the first term, which is simply $[N_X, H]$.
Due to the structure of the Markov generator $H$, this will give a positive contribution for each creation operator of species $X$ in $H$ and a negative contribution for each annihilation operator $a_X$ in $H$. Diagonal terms in $H$ (with the same number of creation and annihilation operators for species $X$) commute with the number operator $N_X$. The commutator $[N_X, H]$ is therefore equal to the off-diagonal terms in $H$, times a prefactor $d$, which counts the difference between the number of creation and annihilation operators of species $X$. 

Additional nested commutators will result in higher powers of $d$, such that the term $ [(N_X)^n, H] $ equals $d^n$ times the off-diagonal terms in $H$ containing creation and/or annihilation operators for species $X$. One can then perform the sum over $n$ in the first exponent of equation \eqref{ZNinter}, which will give a factor $e^{s_x d}$ for each term in $H$, where $d$ is the difference between the number of creation and annihilation operators of species $X$ in this term. A simpler way to obtain the same expression is to replace in $H$ all creation operators $a_X^\dagger$ by $e^{s_x} a_X^\dagger$ and annihilation operators $a_X $ by $e^{-s_x} a_X$, as stated above.

In the case of the profit operators \eqref{profitoperatorsAPP}, we have two observables $\hat{P}_C$ and $\hat{P}_D$. Their moment generating functions are now defined as
\begin{equation}
    M_C(s_c) = \langle e^{\frac{s_c}{t C_0} \hat{N}_{G_c}}\rangle  \,, \qquad  M_D(s_d) = \langle e^{\frac{s_d}{t D_0} \hat{N}_{G_d}} \rangle~~.
\end{equation}
By using the argument above, we may express this as the exponential of tilted generators $\tilde{H}_C(s_c)$ and $\tilde{H}_D(s_d)$
\begin{equation}\label{mgf}
    M_C(s_c) = \left.  e^{\tilde{H}_C \left( \frac{s_c}{t C_0} \right) t} P(0, \vec{z})  \right|_{\vec{z} = \vec{1}}  \,, \qquad  M_D(s_d) = \left.  e^{\tilde{H}_D \left( \frac{s_d}{t D_0} \right) t} P(0, \vec{z})  \right|_{\vec{z} = \vec{1}} \,.
\end{equation}
Here, we have used the fact that the initial distribution $P(0,\vec{z})$ has zero gains, so $e^{s_c \hat{P}_C} P(0,\vec{z}) = P(0,\vec{z})$ and likewise for $\hat{P}_D$. The tilted generators, obtained from the exponential tilting of \eqref{eq:PDgenerator}, are given by
\begin{align}
    \tilde{H}_C \left( \frac{s_c}{t C_0} \right) & =  k_1\left((e^{\frac{s_c}{t C_0}} a^\dagger_{G_c})^{2r} - 1 \right)  (a^\dagger_c)^2 a_c^2 \nonumber\\&+ 2 k_2  \left( (a^\dagger_{G_d})^\tau (e^{\frac{s_c}{t C_0}} a^\dagger_{G_c})^s -1 \right) a^\dagger_c a^\dagger_d a_c a_d   \\&+ k_4 \left( ( a^\dagger_{G_d})^{2p} - 1  \right)(a^\dagger_d)^2 a_d^2~~, \nonumber \\
    \tilde{H}_D \left( \frac{s_d}{t D_0} \right) & =  k_1\left(( a^\dagger_{G_c})^{2r} - 1 \right)  (a^\dagger_c)^2 a_c^2 \nonumber\\&+ 2 k_2  \left( (e^{\frac{s_d}{t D_0}} a^\dagger_{G_d})^\tau (a^\dagger_{G_c})^s -1 \right) a^\dagger_c a^\dagger_d a_c a_d    \\ &+ k_4 \left( (e^{\frac{s_d}{t D_0}} a^\dagger_{G_d})^{2p} - 1  \right)(a^\dagger_d)^2 a_d^2~~. \nonumber
\end{align}
By using these expressions in \eqref{mgf}, together with the initial condition $P(0,\vec{z}) = z_c^{C_0} z_d^{D_0}$ leads to the following expressions for the moment generating functions
\begin{align}
    M_C(s_c)  =   \exp( \lambda_c(s_c) t ) ~~,~~M_D(s_d)  =   \exp( \lambda_d(s_d) t )
\end{align}
where the functions $\lambda_c(s_c)$ and $\lambda_d(s_d)$ are given by
\begin{align} \label{lambdac}
    \lambda_c(s_c) & =  k_1 C_0(C_0-1) \left(e^{2r\, \frac{s_c}{t C_0}}  - 1 \right)  + 2 k_2 C_0 D_0 \left(  e^{s \, \frac{s_c}{t C_0}} -1 \right) \,, \\
    \lambda_d(s_d) & = 2 k_2 C_0 D_0 \left( e^{\tau\, \frac{s_d}{t D_0}} -1 \right) + k_4 D_0(D_0-1) \left(e^{2p\, \frac{s_d}{t D_0} }  - 1 \right)\,,
\end{align}
as was claimed in section \ref{sec:stochasticmodel}. A straightforward application of the same methods for scenarios (b) and (c) leads to the moment generating functions in section \ref{sec:individualgain}.

%\makereferences

\end{document}